\documentclass[aps,showpacs,nofootinbib,superscriptaddress]{revtex4}

\usepackage{graphicx}
\usepackage{epstopdf}
\usepackage{amssymb}
\usepackage{amsmath}
\usepackage{bm}

\newcommand{\g}{\gamma}
\newcommand{\uno}{1 \kern -0.3em {\rm l}}
\newcommand{\nn}{\nonumber}

\newcommand{\sT}{\scriptscriptstyle T}
\newcommand{\half}{\textstyle {\frac{1}{2}}}

\newcommand{\fourth}{\textstyle {\frac{1}{4}}}

\newcommand{\pslash}{\kern 0.2 em p\kern -0.45em /}
\newcommand{\lslash}{\kern 0.2 em l\kern -0.45em /}
\newcommand{\Pslash}{\kern 0.2 em P\kern -0.45em /}
\newcommand{\Sslash}{\kern 0.2 em S\kern -0.45em /}

\begin{document}
\allowdisplaybreaks[2]

\title{Weighted azimuthal asymmetries in a diquark spectator model}

\author{Alessandro Bacchetta}
\email{alessandro.bacchetta@unipv.it}
\affiliation{Dipartimento di Fisica Nucleare e Teorica, Universit\`{a} di 
Pavia, I-27100 Pavia, Italy}
\affiliation{Istituto Nazionale di Fisica Nucleare, Sezione di Pavia, I-27100 
Pavia, Italy}

\author{Marco Radici}
\email{marco.radici@pv.infn.it}
\affiliation{Istituto Nazionale di Fisica Nucleare, Sezione di Pavia, I-27100 
Pavia, Italy}

\author{Francesco Conti}
\email{francesco.conti@pv.infn.it}
\affiliation{Dipartimento di Fisica Nucleare e Teorica, Universit\`{a} di 
Pavia, I-27100 Pavia, Italy}
\affiliation{Istituto Nazionale di Fisica Nucleare, Sezione di Pavia, I-27100 
Pavia, Italy}

\author{Marco Guagnelli}
\email{marco.guagnelli@pv.infn.it}
\affiliation{Istituto Nazionale di Fisica Nucleare, Sezione di Pavia, I-27100 
Pavia, Italy}

\begin{abstract}
We analytically calculate weighted azimuthal asymmetries in semi-inclusive 
lepton-nucleon deep-inelastic scattering and Drell--Yan processes, 
using transverse-momentum-dependent partonic densities obtained in a 
diquark spectator model. We  compare the asymmetries with available 
preliminary experimental data, in particular for the Collins and the Sivers effect. 
We make predictions for other cases of interest in running and planned experiments. 
\end{abstract}

\date{\today}


\maketitle

\section{Introduction}
\label{sec:intro}

Azimuthal asymmetries correspond to cross-section modulations depending
on the azimutal angles involved in the process. Most of the time they are also spin
asymmetries, in the sense that these azimuthal modulations appear with opposite sign 
when the spin of one of the participating particles is reversed. They are essential tools to study 
partonic transverse-momentum distributions (TMDs), defined as probabilities 
to find inside a hadron a parton with longitudinal momentum fraction $x$ and transverse 
momentum $\bm{p}_{\sT}$ with respect to the direction of the parent hadron 
momentum~\cite{Collins:1982uw}. TMDs can be used to construct a
three-dimensional picture of partons  inside hadrons in momentum space and
may be related to the orbital angular momentum of 
partons~\cite{Brodsky:2002cx,Brodsky:2002rv,Burkardt:2003je,Goeke:2006ef,Lu:2006kt,Meissner:2007rx,Qiu:2007ey,Burkardt:2007xm}. 

In a recent paper, we published our results for all leading-twist TMDs in 
the context of a spectator model of the nucleon, using scalar and axial-vector 
diquarks, and further distinguishing between different isospin projections 
($ud$ and $uu$)~\cite{Bacchetta:2008af}. The free 
parameters were fixed by reproducing the parametrizations of 
Ref.~\cite{Chekanov:2002pv} and~\cite{Gluck:2000dy} for the unpolarized and 
helicity distributions at the lowest available scale, respectively, the latter 
being assumed as the model scale. Nonvanishing odd structures with respect to 
na\"ive time-reversal transformations (for brevity, T-odd TMDs), were 
generated by approximating the gauge-link operator with a single gluon-exchange  
interaction, representing the rescattering of the struck quark with the 
spectator diquark at leading order in the strong coupling constant $\alpha_s$. 
Compared to our original publication, in this work we have modified the value we
chose for $\alpha_s$ at the model scale: instead of an {\it ad-hoc}
nonperturbative value, we have computed it using renormalization-group equations at 
leading-order (LO). This change has the effect of resizing all T-odd functions
by a global constant. 

A first test would be to calculate the $\bm{p}_{\sT}$ dependence of
unpolarized cross sections, either in semi-inclusive deep-inelastic scattering 
(SIDIS) or in Drell--Yan hadronic collisions. However, asymmetries 
are usually preferable, because, being defined as ratios of cross-section
combinations, are less sensitive to systematic errors and to theoretical uncertainties as well. 

Two types of azimutal asymmetries are commonly considered: 
weighted and unweighted. From the experimental point of view, unweighted
asymmetries are easier and safer to measure. There is by now a good amount 
of data, mainly on single-spin asymmetries (SSA) in 
SIDIS~\cite{Airapetian:2004tw,Bressan:2009es}.  Experimental data for weighted SSA 
are scarce, with low statistics, and still preliminary~\cite{Seidl:2004dt}. 
However, from the theoretical side unweighted asymmetries are more complex 
because transverse momenta are intertwined in a convolution that can be broken 
only by assuming a specific $\bm{p}_{\sT}$ distribution, typically Gaussian.  Weighted 
asymmetries are preferable: they can be written in a model-independent way in terms of ``collinear" objects --- parton distribution functions (PDFs) or  $\bm{p}_{\sT}$ moments 
of TMDs, and analogously for fragmentation functions~\cite{Mulders:1995dh}.  Most of
the time, this simplifies the theoretical treatment, allows a simpler study of factorization
and scale evolution, and allows more freedom beyond the choice of a Gaussian ansatz 
for TMDs.

In fact, our model TMDs display a $\bm{p}_{\sT}$ distribution which is not 
Gaussian (while also displaying a strong flavor dependence and strong $x-\bm{p}_{\sT}$ correlations)~\cite{Bacchetta:2008af}: as such, it does not allow to analytically
work out unweighted SSA as simple products of terms, and requires a numerical
approach~\cite{futuro} (see also the approximations discussed in 
Ref.~\cite{Boffi:2009sh}). On the contrary, weighted asymmetries in our model can be 
always calculated analytically. 

For what concerns the scale evolution, equations are known only for
the first $\bm{p}_{\sT}$ moment of the Sivers function 
function~\cite{Kang:2008ey,Vogelsang:2009pj,Braun:2009mi}, and 
have a non-diagonal form that makes their treament more difficult than for the standard 
collinear PDFs. In order to include some evolution effects while avoiding
these complications, we compute the moment of the Sivers function at different scales
using only the diagonal part of its evolution equations, and we extend this treatment 
also to the other TMD moments. 

For all the above reasons, we choose to calculate weighted azimuthal (spin)
asymmetries and to compare with the few available data: the double-spin 
asymmetry $A_{LL}$ in SIDIS with longitudinally polarized 
protons~\cite{Airapetian:2004zf}, the Collins and the Sivers effect in 
SIDIS~\cite{Seidl:2004dt}, all measured by the HERMES collaboration. There are 
also $A_{LL}$ data collected by the HERMES and COMPASS collaborations with a 
deuteron target (see for example Ref.~\cite{:2009ci}), but in this paper we will 
consider only measurements directly on a proton target. We will 
show predictions, instead, for other cases of interest in view of running or 
future experiments. For example, at JLab the E06-010 experiment~\cite{e06-010}
is measuring the Collins and the Sivers effects using the 6 GeV 
energy beam hitting on a transversely polarized $^3\mathrm{He}$ (effective neutron) 
target, in order to extract the neutron transversity
distribution. As for Drell--Yan, the fully polarized process with antiprotons 
is planned to be measured at FAIR (GSI), in order to perform a self-consistent 
extraction of the transversity distribution~\cite{Barone:2005pu,Maggiora:2005cr,Efremov:2004qs,Anselmino:2004ki,Bianconi:2005bd};
 while the COMPASS collaboration is planning to measure a SSA using a high 
energetic pion beam on a transversely polarized proton target at 
CERN~\cite{Efremov:2004tp,Bianconi:2006hc,Bianconi:2005yj,Anselmino:2009st}, 
where the T-odd Sivers function could be extracted and its predicted 
non-universal behaviour directly tested~\cite{Collins:2002kn}.

The paper is organized as follows. In Sec.~\ref{sec:ssa}, the kinematics and
the formulae for the weighted azimuthal (spin) asymmetries in SIDIS and 
Drell--Yan are briefly illustrated. In Sec.~\ref{sec:tmd}, the formalism and 
the analytic results for T-even and T-odd TMDs in the spectator diquark model 
of Ref.~\cite{Bacchetta:2008af} are recalled. In Sec.~\ref{sec:out}, the 
updated model T-odd TMDs are compared with the available parametrizations 
and the weighted SSA are compared with the few SIDIS experimental data, making 
predictions for asymmetries in the Drell--Yan process and for other cases of 
interest.


\section{Weighted azimuthal (spin) asymmetries}
\label{sec:ssa}

In the following we will make use of light-cone coordinates. We introduce the 
light-like vectors $n_\pm$ satisfying $n_\pm^2 = 0, \, n_+\cdot n_-=1,$ and we 
describe a generic 4-vector $a = [a^-,a^+,\bm{a}_{\sT}]$, where 
$a^\pm = a\cdot n_\mp$. We will also make use of the transverse tensor
$\epsilon_{\sT}^{ij}= \epsilon^{\mu \nu i j} n_{+\mu} n_{-\nu}$, whose only 
nonvanishing components are $\epsilon_{\sT}^{12}=-\epsilon_{\sT}^{21} = 1$. 

We consider the SIDIS process of a lepton on a (polarized) nucleon, as well as
hadronic collisions leading to Drell--Yan lepton pairs. 


\subsection{SIDIS}
\label{sec:sidis}

In a SIDIS process a (polarized) lepton with momentum $l$ is 
scattered to a final momentum $l'$ by a nucleon target with mass $M$, momentum 
$P$, and spin $S$, leaving also a final hadron semi-inclusively produced with 
mass $M_h$ and momentum $P_h$. The 4-momentum transferred is denoted by 
$q=l-l'$, with $Q^2=-q^2$. We introduce the invariants 
\begin{align}
x &= \frac{Q^2}{2P\cdot q} \approx \frac{p^+}{P^+} \; , &y 
&=\frac{P\cdot q}{P\cdot l} \; , &z &=\frac{P\cdot P_h}{P\cdot q} \approx 
\frac{P_h^-}{k^-} \; , 
\label{eq:lckin}
\end{align}
where $k=p+q$, and $p$ is the momentum of the parton before the scattering. We 
choose a set of frames where $P$ has no transverse components, i.e.,
\begin{equation}
P=\biggl[\frac{M^2}{2 P^+},P^+,\bm{0}\biggr] \; . 
\end{equation} 
Hence, the target polarization 4-vector and the parton momentum can be 
parametrized as 
\begin{align}
S &= \biggl[ -\frac{\lambda_h\, M}{2 P^+}, \frac{\lambda_h\, P^+}{M},
\bm{S}_{\perp} \biggr] \; , \nn \\
p &= \biggl[ \frac{p^2+\bm{p}_{\sT}^2}{2 x P^+},x P^+,\bm{p}_{\sT}\biggr] \; , 
\end{align} 
with $\lambda_h$ the hadron helicity. Analogously, $\lambda_e$ represents the
incoming lepton helicity. 

\begin{figure}[h]
\centering
\includegraphics[width=8.5cm]{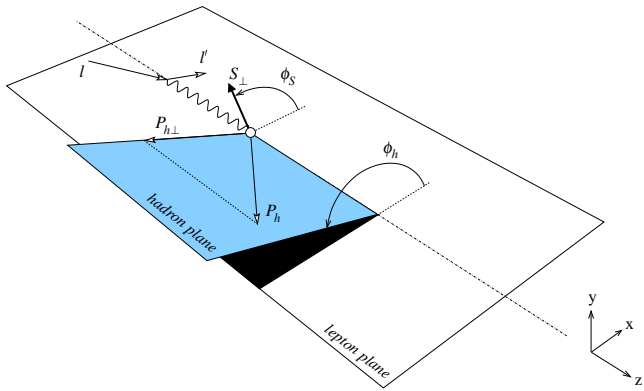}
\caption{The kinematics of the semi-inclusive deep-inelastic scattering.}
\label{fig:sidiskin}
\end{figure}

In single-photon-exchange approximation, the cross section can be 
parametrized in terms of 18 structure functions~\cite{Bacchetta:2006tn}. Here we are  
concerned with the regime $Q^2 \gg \bm{P}^2_{h\perp}, M^2$. An analysis based on TMD factorization reveals that only the following 8 structure functions are present at leading order 
in a $M^2/Q^2$ expansion (twist expansion)~\footnote{Since we are keeping only the 
leading-twist contribution, we have shortened the original notation 
$F_{UU,T},\, F_{UT,T}^{\sin (\phi_h-\phi_S)}$, of Ref.~\cite{Bacchetta:2006tn} to 
$F_{UU},\, F_{UT}^{\sin (\phi_h-\phi_S)}$, since we are not sensitive to longitudinally 
polarized virtual photons.}:
\begin{equation} 
\begin{split} 
\frac{d\sigma}{dx dy dz d\phi_S d\phi_h d\bm{P}^2_{h\perp}} &= 
\frac{\alpha^2}{s}\,\left( 1+\frac{\gamma^2}{2x}\right)\, \biggl\{ 
A(x,y)\,F_{UU} + B(x,y)\,\cos (2\phi_h )\,F_{UU}^{\cos 2\phi_h}   \\
&\quad + \lambda_h\,B(x,y)\,\sin (2\phi_h)\,F_{UL}^{\sin 2\phi_h} + 
\lambda_h\,\lambda_e\,C(x,y)\,F_{LL}   \\ 
&\quad + |\bm{S}_{\perp}|\,\biggl[ 
A(x,y)\,\sin (\phi_h-\phi_S)\,F_{UT}^{\sin (\phi_h-\phi_S)} + 
B(x,y)\,\sin (\phi_h+\phi_S)\,F_{UT}^{\sin (\phi_h+\phi_S)}
\\
& \quad
+ B(x,y)\,\sin (3\phi_h-\phi_S)\,F_{UT}^{\sin (3\phi_h-\phi_S)} \biggr]   
+ |\bm{S}_{\perp}|\,\lambda_e\,C(x,y)\,\cos (\phi_h-\phi_S)\,
F_{LT}^{\cos (\phi_h-\phi_S)} \, \biggr\} \; ,
\label{eq:sidisxsect}
\end{split} 
\end{equation} 
where $\alpha$ is the fine structure constant, $s=Q^2/xy$ is the center-of-mass
energy squared, $\gamma = 2Mx/Q$, $\phi_h,\phi_S,$ are the azimuthal 
orientations of $\bm{P}_h$ and $\bm{S}$ with respect to the scattering plane, 
respectively (see Fig.~\ref{fig:sidiskin}; for their formal definition, see 
Eqs.(2.3)-(2.6) in Ref.~\cite{Bacchetta:2006tn}), and 
\begin{align}
A(x,y) &= \frac{1}{x^2 y^2\, (1+\gamma^2)} \, 
\left( 1-y+\half y^2 + \fourth y^2 \gamma^2 \right) \; ,  \\
B(x,y) &= \frac{1}{x^2 y^2 \, (1+\gamma^2)} \, 
\left( 1-y - \fourth y^2 \gamma^2 \right) \; ,  \\
C(x,y) &= \frac{1}{x^2 y^2 \, \sqrt{1+\gamma^2}} \, y (1-\half y) \; . 
\label{eq:ykin}
\end{align}
Note that the above expressions include mass corrections, related to $\gamma$, and 
the factor $1/(x^2 y^2)$, which must not  be dropped when integrating
separately numerator and denominator of the asymmetries (see Sec.~\ref{sec:out}). 

In Eq.~(\ref{eq:sidisxsect}), each contribution $F_{XY}^W$ depends on 
$x,z,\bm{P}_{h\perp}^2,$ and on $Q^2$; it refers to a lepton with
polarization state $X$ (that can be unpolarized, $U$, or longitudinally
polarized, $L$, with helicity $\lambda_e$), and to a target with polarization
state $Y$ (that can be unpolarized, $U$, longitudinally polarized, $L$, with
helicity $\lambda_h$, and transversely polarized, $T$, with polarization 
$|\bm{S}_{\perp}|$). The superscript $W$ refers to the azimuthal distribution 
of detected hadrons. At leading order in the strong coupling (LO) and leading twist, 
the structure functions read~\cite{Bacchetta:2006tn}
\begin{align}
F_{UU} &= {\cal C} \biggl[ f_1 D_1 \biggr] \; , \nn \\ 
F_{UU}^{\cos 2\phi} &= {\cal C} \biggl[ 
-\frac{2 (\hat{\bm{h}}\cdot \bm{k}_{\sT})\,(\hat{\bm{h}}\cdot \bm{p}_{\sT})-
       \bm{k}_{\sT}\cdot \bm{p}_{\sT}}{M M_h} \, 
h_1^{\perp}\,H_1^{\perp} \biggr] \; , \nn \\
F_{UL}^{\sin 2\phi} &= {\cal C} \biggl[ 
-\frac{2 (\hat{\bm{h}}\cdot \bm{k}_{\sT})\,(\hat{\bm{h}}\cdot \bm{p}_{\sT})-
       \bm{k}_{\sT}\cdot \bm{p}_{\sT}}{M M_h} \, 
h_{1L}^{\perp}\,H_1^{\perp} \biggr] \; , \nn \\
F_{LL} &= {\cal C} \biggl[ g_{1L}\, D_1 \biggr] \; , \nn \\
F_{UT}^{\sin (\phi_h-\phi_S)} &= {\cal C} \biggl[ 
-\frac{\hat{\bm{h}}\cdot \bm{p}_{\sT}}{M}\, f_{1T}^{\perp}\,D_1 \biggr] \; , 
\nn \\ 
F_{UT}^{\sin (\phi_h+\phi_S)} &= {\cal C} \biggl[ 
-\frac{\hat{\bm{h}}\cdot \bm{k}_{\sT}}{M_h}\, h_1\,H_1^{\perp} \biggr] \; , 
\nn \\
F_{UT}^{\sin (3\phi_h-\phi_S)} &= {\cal C} \biggl[ 
\frac{2 (\hat{\bm{h}}\cdot \bm{p}_{\sT})\,(\bm{k}_{\sT}\cdot \bm{p}_{\sT})+
      \bm{p}_{\sT}^2\,(\hat{\bm{h}}\cdot \bm{k}_{\sT})
      -4 (\hat{\bm{h}}\cdot \bm{p}_{\sT})^2\,(\hat{\bm{h}}\cdot \bm{k}_{\sT})}
     {2 M^2 M_h} \, h_{1T}^{\perp}\,H_1^{\perp} \biggr] \; ,  \nn \\
F_{LT}^{\cos (\phi_h-\phi_S)} &= {\cal C} \biggl[ 
\frac{\hat{\bm{h}}\cdot \bm{p}_{\sT}}{M}\, g_{1T}\, D_1 \biggr] \; , 
\label{eq:sidisF}
\end{align}
where $\hat{\bm{h}} = \bm{P}_{h\perp} / |\bm{P}_{h\perp}|$, and the 
convolution is defined by 
\begin{equation}
{\cal C}\biggl[ w\,f\,D \biggr] = x \sum_a e_a^2 \int d\bm{p}_{\sT}
d\bm{k}_{\sT} \, \delta^{(2)}(\bm{p}_{\sT}-\bm{k}_{\sT}-\bm{P}_{h\perp}/z)\, 
w(\bm{p}_{\sT}, \bm{k}_{\sT})\, f^a(x,\bm{p}_{\sT}^2)\, D^a (z,\bm{k}_{\sT}^2) 
\; , 
\label{eq:convolution}
\end{equation}
with the sum running over all parton and antiparton flavors $a$. 

The different structure functions can be extracted thanks to their specific dependence 
on the azimuthal angles. Upon integration in $d\bm{P}_{h\perp}^2$, only the 
convolutions in $F_{UU}$ and $F_{LL}$ can be solved analytically, because of the 
simple weight $w=1$. In this case, the well known double-spin asymmetry can be 
formed,  
\begin{equation}
A_{LL} = \frac{C(x,y)}{A(x,y)}\, 
\frac{\int d\bm{P}_{h\perp}^2\, F_{LL}}{\int d\bm{P}_{h\perp}^2\, F_{UU}} = 
\frac{C(x,y)}{A(x,y)}\, 
\frac{\sum_a e_a^2 \, x g_1^a(x)\,D_1^a(z)}
     {\sum_a e_a^2\, x f_1^a(x)\,D_1^a(z)} \equiv  
\frac{C(x,y)}{A(x,y)}\, A_1  \; , 
\label{eq:all}
\end{equation}
giving access to the helicity distribution 
$g_1^a(x) = \int d\bm{p}_{\sT}\,g_{1L}^a(x,\bm{p}_{\sT}^2)$. This asymmetry is analyzed in
Sec.~\ref{sec:sidis-out}.

In the other cases, an analytical solution can be achieved only with a
suitable model dependence upon 
$\bm{p}_{\sT}$ and $\bm{k}_{\sT}$ of the distribution and fragmentation 
functions, respectively~\cite{Mulders:1995dh}. 
The most widely used Ansatz is a Gaussian distribution in $\bm{p}_{\sT}$ 
and $\bm{k}_{\sT}$ (usually also assumed to be flavor independent and 
independent of $x$ and $z$). It allows an analytic calculation of all convolutions 
in Eq.~(\ref{eq:convolution}). However, theoretical calculations indicate that perturbative corrections generate non-Gaussian tails of the TMDs when 
$\bm{P}_{h\perp}^2\gg M^2$~\cite{Bacchetta:2008xw} 
and available experimental data poorly constrain the $\bm{p}_{\sT}$  
dependence of partonic densities in general, leaving room for Ans\"atze 
different from the Gaussian one. Actually, our model displays a $\bm{p}_{\sT}$ 
distribution which is not factorized, has a strong flavor dependence, and 
above all is not Gaussian~\cite{Bacchetta:2008af}. As such, it does not allow 
to analytically calculate all the convolutions in Eq.~(\ref{eq:convolution}). 
In this paper we will consider only properly 
$\bm{P}_{h\perp}-$weighted asymmetries, which break the convolutions 
in a model-independent way and result  in factorized expressions such as 
the one in Eq.~(\ref{eq:all}). We leave the computation of unweighted asymmetries 
to a future work, where we will consider a direct (numerical) calculation of all 
the convolutions needed in Eq.~(\ref{eq:sidisF})~\cite{futuro}. 

Weighted asymmetries are defined as~\cite{Boer:1997nt}
\begin{equation}
A_{XY}^W(x,y,z) \propto \frac{\langle W \rangle_{XY}}{\langle 1 \rangle_{UU}} 
\equiv \frac{\int d\phi_S d\phi_h d\bm{P}^2_{h\perp}\,W\, d\sigma_{XY}}
            {\int d\phi_S d\phi_h d\bm{P}^2_{h\perp}\,d\sigma_{UU}} \; , 
\label{eq:wasymmdef}
\end{equation}
where $d\sigma_{XY}$ refers to the contribution in Eq.~(\ref{eq:sidisxsect}) 
for the lepton probe with polarization $X$ and nucleon target with 
polarization $Y$. Typically, the weight $W$ can be function of 
$\phi_S, \phi_h,$ and of suitable powers of $Q_{\sT} = |\bm{P}_{h\perp}| / z$. 

For SIDIS, the following weighted 
azimuthal asymmetries are relevant and theoretically clean~\cite{Boer:1997nt}:
\begin{align}
A_{UT}^{Q_T\,\sin (\phi_h+\phi_S)} &= 2\,
\frac{\Bigl\langle \frac{Q_T}{M_h}\, \sin (\phi_h+\phi_S) \Bigr\rangle_{UT}}
     {\langle 1 \rangle_{UU}} = 2\, \frac{B(x,y)}{A(x,y)} \, 
\frac{\sum_a e_a^2 \, x h_1^a(x)\,H_1^{\perp (1) a}(z)}
     {\sum_a e_a^2\, x f_1^a(x)\,D_1^a(z)} \; , \label{eq:Autsinph+S} \\
A_{UT}^{Q_T\,\sin (\phi_h-\phi_S)} &= 2\,
\frac{\Bigl\langle \frac{Q_T}{M}\, \sin (\phi_h-\phi_S) \Bigr\rangle_{UT}}
     {\langle 1 \rangle_{UU}} = -2\, \frac{A(x,y)}{A(x,y)} \, 
\frac{\sum_a e_a^2 \, x f_{1T}^{\perp (1) a}(x)\,D_1^a(z)}
     {\sum_a e_a^2\, x f_1^a(x)\,D_1^a(z)} \; , \label{eq:Autsinph-S} \\
A_{LT}^{Q_T\,\cos (\phi_h-\phi_S)} &= 2\,
\frac{\Bigl\langle \frac{Q_T}{M}\, \cos (\phi_h-\phi_S) \Bigr\rangle_{UT}}
     {\langle 1 \rangle_{UU}} = 2\, \frac{C(x,y)}{A(x,y)} \, 
\frac{\sum_a e_a^2 \, x g_{1T}^{(1) a}(x)\,D_1^a(z)}
     {\sum_a e_a^2\, x f_1^a(x)\,D_1^a(z)} \; , \label{eq:Altcosph-S} 
\end{align}
where 
\begin{equation}
f^{(n)}(x) = \int d\bm{p}_{\sT} \left( \frac{\bm{p}_{\sT}^2}{2M^2} \right)^n
\, f(x,\bm{p}_{\sT}^2)  
\label{eq:ptmoment}
\end{equation}
is the $n$-th $\bm{p}_{\sT}$-moment of a parton distribution $f$ (and 
analogously for a fragmentation function). 

In every parton density 
above (or $\bm{p}_{\sT}$ moment of it), the dependence on the hard scale $Q^2$
is understood. Evolution effects with running $Q^2$ are well known for 
$f_1,\, g_1,\, h_1,$ and $D_1$.  From the discussion in Sec. 7.3 of 
Ref.~\cite{Bacchetta:2008xw} and the explicit study of Ref.~\cite{Vogelsang:2009pj}, it 
should be evident why we can hope to study the weighted 
asymmetries~(\ref{eq:Autsinph+S} - \ref{eq:Altcosph-S}) in a way similar to the 
double-spin asymmetry of Eq.~\eqref{eq:all}, including the scale dependence 
of the involved functions. Explicit calculations are available only for the evolution 
equations of the Sivers function 
$f_{1T}^{\perp (1)}$~\cite{Kang:2008ey,Vogelsang:2009pj,Braun:2009mi}. 
They are nondiagonal and thus considerably more complex than those of twist-2 collinear 
PDFs. However, their diagonal part is identical to the evolution equations for $f_1$ 
and should be dominant at high $x$~\cite{Vogelsang:2009pj}. Calculations of the 
real perturbative contributions in $g_{1T}$~\cite{Zhou:2009jm} suggest that the evolution equations for this function will have a diagonal part identical to that of $g_1$. 
In summary, we approximately implement the effect of scale evolution in the
weighted asymmetries~(\ref{eq:Autsinph+S} - \ref{eq:Altcosph-S}) 
by evolving $f_{1T}^{\perp (1)}(x)$ and $g_{1T}^{(1)}(x)$ using the same LO evolution of
 $f_{1}(x)$ and $g_{1}(x)$ (see, e.g., Ref.~\cite{Barone:2001sp}). 
 For $H_1^{\perp (1)}(z)$, we used the same LO evolution as the transversity 
 fragmentation function $H_1(z)$ (see, e.g., Ref.~\cite{Stratmann:2001pt}).  
The asymmetries~(\ref{eq:Autsinph+S} - \ref{eq:Altcosph-S}) are considered in 
Sec.~\ref{sec:sidis-out}.

In our analysis, we will not take into consideration the following asymmetries
\begin{align}
A_{UU}^{Q_T^2\,\cos 2\phi_h} &= 2\,
\frac{\Bigl\langle \frac{Q_T^2}{4M M_h}\, \cos 2\phi_h \Bigr\rangle_{UU}}
     {\langle 1 \rangle_{UU}} \stackrel{\rm TMD}{=} 
2\, \frac{B(x,y)}{A(x,y)} \, 
\frac{\sum_a e_a^2 \, x h_1^{\perp (1) a}(x)\,H_1^{\perp (1) a}(z)}
     {\sum_a e_a^2\, x f_1^a(x)\,D_1^a(z)} \; , \label{eq:Auucos2ph} \\
A_{UL}^{Q_T^2\,\sin 2\phi_h} &= 2\,
\frac{\Bigl\langle \frac{Q_T^2}{4M M_h}\, \sin 2\phi_h \Bigr\rangle_{UL}}
     {\langle 1 \rangle_{UU}}\stackrel{\rm TMD}{=}  2\, \frac{B(x,y)}{A(x,y)} \, 
\frac{\sum_a e_a^2 \, x h_{1L}^{\perp (1) a}(x)\,H_1^{\perp (1) a}(z)}
     {\sum_a e_a^2\, x f_1^a(x)\,D_1^a(z)} \; , \label{eq:Aulsin2ph} \\
A_{UT}^{Q_T^3\,\sin (3\phi_h-\phi_S)} &= 2\,
\frac{\Bigl\langle \frac{Q_T^3}{6M^2 M_h}\, \sin (3\phi_h-\phi_S) \Bigr\rangle_{UT}}
     {\langle 1 \rangle_{UU}} \stackrel{\rm TMD}{=}  2\, \frac{B(x,y)}{A(x,y)} \, 
\frac{\sum_a e_a^2 \, x h_{1T}^{\perp (2) a}(x)\,H_1^{\perp (1) a}(z)}
     {\sum_a e_a^2\, x f_1^a(x)\,D_1^a(z)} \; , \label{eq:Autsin3ph-S} \\
\end{align}
because the higher powers of $Q_{\sT}$ in the weight emphasize the 
high-$\bm{p}_{\sT}$ components of the asymmetry, that are
dominated by perturbative QCD corrections~\cite{Bacchetta:2008xw} and do not
give enough information about TMDs.  In the above equations, we nevertheless gave 
the explicit expressions for these weighted asymmetries in the TMD framework, 
since it might still be possible to isolate their TMD component by means of
differences, or ratios between different measurements, that cancel the
perturbative contributions. We will not discuss these possibilities in the
present work.


\subsection{Drell--Yan}
\label{sec:dy}

In a Drell--Yan process, a pair with individual momenta $k_1$ and $k_2$, formed 
by a lepton and its antilepton partner, is produced from the collision of two 
hadrons with momentum $P_i$, mass $M_i$, and spin $S_i$, with $i=1,2$. The cm 
square energy available is $s=(P_1+P_2)^2$; the momentum transfer is now 
time-like and gives the invariant mass of the pair, i.e. 
\begin{equation}
q^2 \equiv M^2 = (k_1 + k_2)^2 = (p_1+p_2)^2 \; , 
\label{eq:tlq}
\end{equation}
where $p_i$ are the momenta of the annihilating partons, with $i=1,2$. If 
$P_1^+$ and $P_2^-$ are the dominant light-cone components of hadron momenta 
in the regime where the factorization theorem holds~\cite{Collins:1984kg}, we 
can define the following invariants: 
\begin{align}
x_1 &= \frac{q^2}{2P_1\cdot q} \approx \frac{p_1^+}{P_1^+} \; , \qquad x_2 = 
\frac{q^2}{2P_2\cdot q} \approx \frac{p_2^-}{P_2^-} \; , \nn \\
y &= \frac{k_1\cdot P_1}{q\cdot P_1} \approx \frac{k_1^-}{q^-}\; , \quad \tau = 
\frac{M^2}{s} \; , \quad  x_F = x_1 - x_2 \; . 
\label{eq:dyinvariants}
\end{align}

We choose $x_1$ as the momentum fraction of the parton in the beam; namely, 
\begin{align}
x_1 &= \half \, \left( x_F + \sqrt{x_F^2+4\tau}\right) = \sqrt{\tau}\,e^y \; , 
\nn \\
x_2 &= \half \, \left( -x_F + \sqrt{x_F^2+4\tau}\right) = \sqrt{\tau}\,e^{-y} 
\; . 
\label{eq:dykinvar}
\end{align}
In the following, we will plot SSA as functions of $x_F$ and/or $y$ at a given 
$\tau$: this choice will probe very different regions in $x_1$ and $x_2$, 
where the model parton distributions involved behave very differently. And 
this feature will show up clearly in the asymmetry. 

\begin{figure}[h]
\centering
\includegraphics[width=7cm]{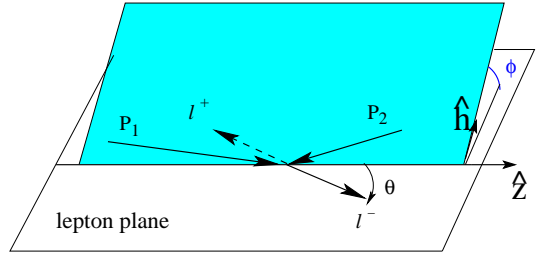}
\caption{The Collins-Soper frame.}
\label{fig:dyframe}
\end{figure}

Because of momentum conservation, the 4-momentum transfer can be parametrized 
as
\begin{equation}
q = \biggl[ x_2 P_2^- , x_1 P_1^+ , \bm{p}_{1\sT} + \bm{p}_{2\sT} \biggr] \; .
\label{eq:tlq2}
\end{equation}
If the transverse momentum of the final lepton pair is 
$\bm{q}_{\sT} = \bm{p}_{1\sT}+\bm{p}_{2\sT} \neq 0$, the directions of the two 
annihilating partons are not known. Hence, it is convenient to select the socalled 
Collins-Soper frame~\cite{Collins:1977iv} described in Fig.~\ref{fig:dyframe}. 
The final lepton pair is detected in the solid angle $(\theta, \phi )$, where, 
in particular, $\phi$ (and all other azimuthal angles) is measured in a plane 
perpendicular to the indicated lepton plane but containing 
$\hat{\bm{h}} = \bm{q}_{\sT} / |\bm{q}_{\sT}| \equiv \bm{q}_{\sT} / q_{\sT} $. 

The fully differential LO cross section can contain many contributions 
according to the polarization state of the two colliding 
hadrons~\cite{Arnold:2008kf}. Having in mind the phenomenology that could be 
explored at GSI-FAIR and CERN, or at J-PARC and RHIC, we limit the physics 
case to the most relevant combinations. The master channel is the 
$\bar{p}^\uparrow p^\uparrow \rightarrow l^+ l^- X$ 
reaction~\cite{Efremov:2004qs,Barone:2005pu,Maggiora:2005cr}, whose polarized 
$q_{\sT}-$integrated cross section is simply~\cite{Tangerman:1995eh}
\begin{equation}
\frac{d \tilde{\sigma}^{\uparrow \uparrow}}{d\Omega dx_1 dx_2} = 
\frac{\alpha^2}{3 q^2}\,|\bm{S}_{1\sT}|\,|\bm{S}_{2\sT}|\, \fourth 
\sin^2\theta \, \cos (2\phi -\phi_{S_1} -\phi_{S_2})\, \sum_a e_a^2 \, 
h_1^{\overline a}(x_1)\,h_1^a(x_2) \; , 
\label{eq:ttxsect}
\end{equation}
where $d\Omega = \sin \theta d\theta d\phi$ and $|\bm{S}_{i\sT}|,\,\phi_{S_i}$
are the transverse polarization and the azimuthal angle of its vector for 
hadron $i$. Combining Eq.~(\ref{eq:ttxsect}) with the unpolarized part, we can 
form the well known double-spin asymmetry 
\begin{align}
\tilde{A}_{TT} &= 
\frac{d\tilde{\sigma}^{\uparrow \uparrow}-d\tilde{\sigma}^{\uparrow \downarrow}}
     {d\tilde{\sigma}^{\uparrow \uparrow}+d\tilde{\sigma}^{\uparrow \downarrow}}
\nn \\
&= |\bm{S}_{1\sT}|\,|\bm{S}_{2\sT}|\, \frac{\sin^2\theta}{1+\cos^2\theta}\,
\cos (2\phi -\phi_{S_1} -\phi_{S_2})\, 
\frac{\sum_a e_a^2 \,x_1 h_1^{\overline a}(x_1)\,x_2 h_1^a(x_2)}
       {\sum_a e_a^2 \,x_1 f_1^{\overline a}(x_1)\,x_2 f_1^a(x_2)} \nn \\
&\equiv |\bm{S}_{1\sT}|\,|\bm{S}_{2\sT}|\, \tilde{a}_{TT} \, 
\frac{\sum_a e_a^2 \,x_1 h_1^{\overline a}(x_1)\,x_2 h_1^a(x_2)}
       {\sum_a e_a^2 \,x_1 f_1^{\overline a}(x_1)\,x_2 f_1^a(x_2)} \; , 
\label{eq:dyatt}
\end{align}
which gives direct access to (the valence part of) the transversity 
distribution $h_1$. Monte Carlo simulations about the feasibility of this 
measurement were presented in Ref.~\cite{Bianconi:2005bd}. Predictions based 
on our diquark spectator model will be shown in Sec.~\ref{sec:dy-out}. 

When in the collision only one hadron can be transversely polarized, the fully
differential leading-twist cross section reads~\cite{Boer:1999mm,Arnold:2008kf}
\begin{align}
\frac{d\tilde{\sigma}}{d\Omega dx_1 dx_2 d\bm{q}_{\sT}} &= 
\frac{\alpha^2}{3q^2}\,\Biggl\{ \tilde{A}(y) \, \tilde{F}_{UU} + 
\tilde{B}(y)\,\cos (2\phi ) \, \tilde{F}_{UU}^{\cos 2\phi} \nn \\
& \hspace{-2cm} + |\bm{S}_{2\sT}|\, \biggl[ \tilde{A}(y)\, 
\sin (\phi -\phi_{S_2}) \, \tilde{F}_{UT}^{\sin (\phi -\phi_{S_2})} -
\tilde{B}(y) \,\sin (\phi +\phi_{S_2})\, 
\tilde{F}_{UT}^{\sin (\phi +\phi_{S_2})} - \tilde{B}(y) \, 
\sin (3\phi -\phi_{S_2})\, \tilde{F}_{UT}^{\sin (3\phi -\phi_{S_2})} \biggr] 
\, \Biggr\} \; , 
\label{eq:dyxsect}
\end{align}
where  
\begin{align}
\tilde{A}(y) = \left( \half - y + y^2 \right) \, \stackrel{\mbox{cm}}{=}\, 
\fourth \left( 1 + \cos^2 \theta \right) &\mbox{\hspace{2cm}} 
\tilde{B}(y) = y (1-y) \, \stackrel{\mbox{cm}}{=}\,\fourth \, \sin^2 \theta 
\; . \label{eq:dylepton}
\end{align}

The structure functions $\tilde{F}_{XY}^W$ depend on $x_1,x_2, q_{\sT}$.
They read~\cite{Arnold:2008kf}
\begin{align}
\tilde{F}_{UU} &= \tilde{\cal C}\biggl[ 1\, f_1\, f_1 \biggr] \; ,  \nn \\
\tilde{F}_{UU}^{\cos 2\phi} &= \tilde{\cal C}\biggl[ 2 
\frac{\hat{\bm{h}}\cdot \bm{p}_{1\sT} \, \hat{\bm{h}} \cdot \bm{p}_{2\sT} - 
      \bm{p}_{1\sT} \cdot \bm{p}_{2\sT}}{M_1 M_2} \, h_1^{\perp}\, h_1^{\perp} 
\biggr] \; , \nn \\
\tilde{F}_{UT}^{\sin (\phi -\phi_{S_2})} &= \tilde{\cal C}\biggl[ 
\frac{\hat{\bm{h}}\cdot \bm{p}_{2\sT}}{M_2} \, f_1 \, f_{1\sT}^{\perp} \biggr] 
\; , \nn \\
\tilde{F}_{UT}^{\sin (\phi +\phi_{S_2})} &= \tilde{\cal C}\biggl[ 
\frac{\hat{\bm{h}}\cdot \bm{p}_{1\sT}}{M_1} \, h_1^{\perp} \, h_1 \biggr] 
\; , \nn \\
\tilde{F}_{UT}^{\sin (3\phi -\phi_{S_2})} &= \tilde{\cal C}\biggl[ 
\frac{4 \hat{\bm{h}}\cdot \bm{p}_{1\sT} \, (\hat{\bm{h}}\cdot \bm{p}_{2\sT})^2 
      - 2 \hat{\bm{h}} \cdot \bm{p}_{2\sT} \, \bm{p}_{1\sT}\cdot \bm{p}_{2\sT} 
      - \hat{\bm{h}}\cdot \bm{p}_{1\sT} \, \bm{p}_{2\sT}^2}{2 M_1\,M_2^2}\, 
      h_1^{\perp}\, h_{1\sT}^{\perp} \biggr] \; . 
\label{eq:dyF}
\end{align}
The convolution $\tilde{\cal C}$ is defined as 
\begin{equation}
\tilde{\cal C}\biggl[ w\,f\,g \biggr] = x_1 x_2 \sum_a e_a^2 \int 
d\bm{p}_{1\sT} d\bm{p}_{2\sT} \, \delta^{(2)}(\bm{p}_{1\sT}+\bm{p}_{2\sT}-
\bm{q}_{\sT})\, w(\bm{p}_{1\sT},\bm{p}_{2\sT})\, \biggl[ 
f^{\overline a}(x_1,\bm{p}_{1\sT}^2)\, g^a(x_2,\bm{p}_{2\sT}^2)  \biggr] \; . 
\label{eq:convolution2}
\end{equation}

As in the SIDIS case, only the convolution in $\tilde{F}_{UU}$ can be solved
analytically upon integration in $d\bm{q}^2_{\sT}$, because of the simple
weight $w=1$. Again, instead of introducing a suitable model dependence upon 
transverse momenta, we follow the strategy of considering properly 
$q_{\sT}-$weighted asymmetries, 
 now defined as
\begin{equation}
\tilde{A}_{XY}^W(x_1,x_2,y) \propto 
\frac{\langle W \rangle_{XY}}{\langle 1 \rangle_{UU}} \equiv 
\frac{\int d\phi_{S_2} d\phi d\bm{q}_{\sT}^2\,W\, d\tilde{\sigma}_{XY}}
     {\int d\phi_{S_2} d\phi d\bm{q}_{\sT}^2\,d\tilde{\sigma}_{UU}} \; , 
\label{eq:wasymmdef2}
\end{equation}
where $d\tilde{\sigma}_{XY}$ refers to the contribution in 
Eq.~(\ref{eq:dyxsect}) for the hadrons 1 and 2 with polarizations $X$ and $Y$,
respectively. Typically, the weight $W$ can be function of $\phi_{S_2}, \phi,$ 
and of suitable powers of $q_{\sT}$. 

We consider the following weighted asymmetries: 
\begin{align}
\tilde{A}_{UT}^{q_T\,\sin (\phi -\phi_{S_2})} &= 2\,
\frac{\Bigl\langle \frac{q_T}{M_2}\, \sin (\phi -\phi_{S_2}) \Bigr\rangle_{UT}}
     {\langle 1 \rangle_{UU}} = 2\, \frac{\tilde{A}(y)}{\tilde{A}(y)} \, 
\frac{\sum_a e_a^2 \, x_1 f_1^{\overline a}(x_1)\,x_2 f_{1T}^{\perp (1) a}(x_2)}
       {\sum_a e_a^2\, x_1 f_1^{\overline a}(x_1)\,x_2 f_1^a(x_2)} \; , 
\label{eq:dyAutsinph-S} \\
\tilde{A}_{UT}^{q_T\,\sin (\phi +\phi_{S_2})} &= 2\,
\frac{\Bigl\langle \frac{q_T}{M_1}\, \sin (\phi +\phi_{S_2}) \Bigr\rangle_{UT}}
     {\langle 1 \rangle_{UU}} = -2\, \frac{\tilde{B}(y)}{\tilde{A}(y)} \, 
\frac{\sum_a e_a^2 \, x_1 h_1^{\perp (1) {\overline a}}(x_1)\,x_2 h_1^a(x_2)}
        {\sum_a e_a^2\, x_1 f_1^{\overline a}(x_1)\,x_2 f_1^a(x_2)} \; , 
\label{eq:dyAutsinph+S}
\end{align}
For the same considerations expressed in the case of SIDIS, we refrain
ourselves from considering the following asymmetries
\begin{align}
\tilde{A}_{UU}^{q_T^2\,\cos 2\phi} &= 2\,
\frac{\Bigl\langle \frac{q_T^2}{4M_1 M_2}\, \cos (2\phi ) \Bigr\rangle_{UU}}
     {\langle 1 \rangle_{UU}} \stackrel{\rm TMD}{=} 2\, 
\frac{\sum_a e_a^2 \, x_1 h_1^{\perp (1) {\overline a}}(x_1)\,x_2 h_1^{\perp (1) a}(x_2)}
        {\sum_a e_a^2\, x_1 f_1^{\overline a}(x_1)\,x_2 f_1^a(x_2)} \; , 
\label{eq:dyAuucos2ph} \\
\tilde{A}_{UT}^{q_T^3\,\sin (3\phi -\phi_{S_2})} &= 2\,
\frac{\Bigl\langle \frac{q_T^3}{6M_1 M_2^2}\, \sin (3\phi -\phi_{S_2}) \Bigr\rangle_{UT}}
     {\langle 1 \rangle_{UU}} \stackrel{\rm TMD}{=} -2\, \frac{\tilde{B}(y)}{\tilde{A}(y)} \, 
\frac{\sum_a e_a^2 \, x_1 h_1^{\perp (1) {\overline a}}(x_1)\,x_2 h_{1T}^{\perp (2) a}(x_2)}
        {\sum_a e_a^2\, x_1 f_1^{\overline a}(x_1)\,x_2 f_1^a(x_2)} \; , 
\label{eq:dyAutsin3ph-S} 
\end{align}
where the right-hand-side represents only the TMD contribution to the
asymmetry, which is suppressed with respect to perturbative contributions. 

Monte Carlo simulations about the feasibility of measurements of the 
asymmetries~(\ref{eq:dyAutsinph-S}) and (\ref{eq:dyAutsinph+S}) at GSI-FAIR, 
COMPASS, and RHIC, were presented in 
Refs.~\cite{Bianconi:2004wu,Bianconi:2006hc,Bianconi:2005yj,Radici:2007vc}. 
Predictions based on our diquark spectator model are displayed in 
Sec.~\ref{sec:dy-out}. Similarly, while in Eqs.~(\ref{eq:dyAutsinph-S},\ref{eq:dyAutsinph+S}) 
the dependence on the hard scale $Q^2$ is understood, evolution effects with 
running $Q^2$ are included in the same way as discussed at the end of the
previous Sec.~\ref{sec:sidis}. The chiral-odd function $h_1^{\perp (1)}$ is
assumed to follow the same LO evolution equations as the chiral-odd function $h_1$.



\section{T-even and T-odd partonic densities in a diquark spectator model}
\label{sec:tmd}

The TMD showing up in Eqs.~(\ref{eq:Autsinph+S}-\ref{eq:Altcosph-S}) and
(\ref{eq:dyAutsinph-S},\ref{eq:dyAutsinph+S}) can be calculated analytically in
the spectator diquark model. They can be extracted from the basic quantity named
quark-quark correlator~\cite{Bacchetta:2006tn}, i.e. 
\begin{equation} 
\Phi (x,\bm{p}_{\sT},S)= \int \frac{d\xi^- d\bm{\xi}_{\sT}}{(2\pi)^3}\; 
       e^{i p \cdot \xi}\,\langle P,S|\bar{\psi}(0)\,{\cal U}_{[0,\xi]}\,
       \psi(\xi)|P,S \rangle \Big|_{\xi^+=0} \; ,
\label{eq:Phi-tree}
\end{equation}
where 
\begin{equation} 
U_{[0,\xi]} = {\cal P}\, e^{-ig \int_0^\xi dw \cdot A(w)}
\label{eq:link}
\end{equation}
is the socalled gauge link operator connecting the two different space-time 
points $0$ and $\xi$ by a specific path followed by the gluon field 
$A$, which couples to the quark field $\psi$ through the coupling $g$. The 
gauge link ensures that the matrix element of Eq.~(\ref{eq:Phi-tree}) is 
color-gauge invariant.

\begin{figure}[h]
\begin{center}
\includegraphics[width=6cm]{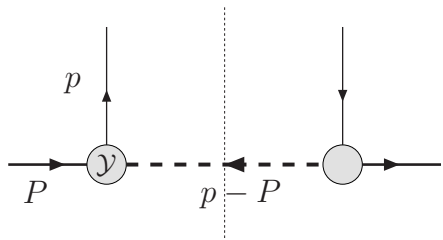}
\end{center}
\caption{Tree-level cut diagram for the calculation of T-even 
leading-twist parton densities. The dashed line indicates both 
scalar and axial-vector diquarks.}
\label{fig:sidis}
\end{figure}

The correlator~(\ref{eq:Phi-tree}) can be evaluated in the spectator 
approximation~\cite{Jakob:1997wg,Bacchetta:2008af}: a completeness relation is 
inserted and at tree-level the sum over final states is truncated to a single 
on-shell spectator state with mass $M_X$ (see Fig.~\ref{fig:sidis}),  
\begin{equation} 
\Phi(x,\bm{p}_{\sT},S) \sim \frac{1}{(2\pi)^3}\,\frac{1}{2(1-x)P^+}\, 
\overline{\mathcal{M}}^{(0)}(S)\, \mathcal{M}^{(0)}(S) 
\Big\vert_{p^2=\tau (x,\bm{p}_{_T})}\; , 
\label{eq:Phi-tree-spect}
\end{equation}
where the on-shell condition $(P-p)^2=M_X^2$ for the spectator implies for the 
quark the off-shell condition
\begin{align} 
p^2 \equiv \tau (x,\bm{p}_{\sT}) &=-\frac{\bm{p}_{\sT}^2+L_X^2(m^2)}{1-x}+m^2 
\; , &L_X^2(m^2)&=x M_X^2 + (1-x) m^2 - x (1-x) M^2 \; .
\label{eq:offshell} 
\end{align} 

We assume the spectator to be point-like, with the quantum numbers of a
diquark. Hence, the proton can couple to a quark and to a spectator diquark 
with spin 0 (scalar $X=s$) or spin 1 (axial-vector $X=a$), as well as with 
isospin projection 0 ($ud$-like system) or 1 ($uu$-like system). 
Therefore, the tree-level ``scattering amplitude'' $\mathcal{M}^{(0)}$ 
is given by (see Fig.~\ref{fig:sidis})~\cite{Bacchetta:2008af} 
\begin{equation} 
\mathcal{M}^{(0)}(S) = \langle P-p |\psi(0) |P,S\rangle = \
  \begin{cases}
  \displaystyle{\frac{i}{\pslash-m}}\, {\cal Y}_s \, U(P,S)& 
        \text{scalar diquark,} \\ 
  \displaystyle{\frac{i}{\pslash-m}}\, \varepsilon^*_{\mu}(P-p,\lambda_a)\,
        {\cal Y}^{\mu}_a \, U(P,S)& \text{axial-vector diquark,} 
  \end{cases}
\label{eq:m-tree}
\end{equation}
where the nucleon-quark-diquark vertex can have the form 
\begin{align} 
{\cal Y}_s &= i g_s(p^2)\, \uno \quad \text{scalar}\; , &{\cal Y}^{\mu}_a &= 
i \frac{g_a(p^2)}{\sqrt{2}}\,\g^\mu \, \g_5\quad \text{axial-vector} \; ,
\label{eq:tree-vertices}
\end{align} 
with $g_X(p^2)$ a suitable form factor. The $\varepsilon_\mu(P-p,\lambda_a)$ is 
the 4-vector polarization of the spin-1 axial-vector diquark with momentum 
$P-p$ and helicity states $\lambda_a$. In Ref.~\cite{Bacchetta:2008af}, several 
choices were analyzed for the diquark polarization sum 
$d^{\mu \nu}= \sum_{\lambda_a} \varepsilon^{\ast \mu}_{(\lambda_a)} \, 
\varepsilon^\nu_{(\lambda_a)}$ and for the form factor $g_X(p^2)$, with all
analytic formulae listed in the appendices. Numerical results were presented
only for $d^{\mu \nu}$ propagating transverse diquark polarizations, and
for the so-called ``dipolar" choice
\begin{equation} 
g_X(p^2) = g_X\,\frac{p^2-m^2}{{|p^2-\Lambda_X^2|}^2} = g_X\,
\frac{(p^2-m^2)\,(1-x)^2}{\left( \bm{p}_{\sT}^2+L_X^2(\Lambda_X^2)\right)^2} 
\; . 
\label{eq:ffdip}
\end{equation}
Here, we keep the same choices and in the next subsections we list the results
for the T-even and T-odd leading-twist TMDs.


\subsection{T-even}
\label{sec:teven}

The prototype of TMDs is the unpolarized parton distribution, which can be
defined in the following way:
\begin{equation} 
\begin{split}  
f_1(x,\bm{p}_{\sT}) &= \frac{1}{4}\,\mathrm{Tr}\left[ \left( 
\Phi(x,\bm{p}_{\sT},S) + \Phi(x,\bm{p}_{\sT},-S) \right) \, \g^+ \right] 
+ \mathrm{h.c.}  \\
&= \frac{1}{4}\,\frac{1}{(2\pi)^3}\,\frac{1}{2(1-x)P^+}\, \mathrm{Tr}\left[ 
\left( \overline{\mathcal{M}}^{(0)}(S)\, \mathcal{M}^{(0)}(S) + 
\overline{\mathcal{M}}^{(0)}(-S)\, \mathcal{M}^{(0)}(-S) \right) \, \g^+ 
\right] + \mathrm{h.c.} \; , 
\label{eq:f1}
\end{split} 
\end{equation}
and can be computed 
by inserting in $\mathcal{M}^{(0)}$ of Eq.~(\ref{eq:m-tree}) the 
rules~(\ref{eq:tree-vertices}) for the nucleon-quark-diquark vertex, the 
dipolar form factor of Eq.~(\ref{eq:ffdip}), and the diquark polarization sum.

To express in a condensed way all the other TMDs, it is convenient to
 introduce the two combinations
\begin{align} 
f_1^{+ q(s)} (x,\bm{p}_{\sT}^2) &\equiv \frac{1}{2}\Bigl(f_1^{q(s)}
(x,\bm{p}_{\sT}^2)  +g_{1L}^{q(s)} (x,\bm{p}_{\sT}^2)\Bigr)
=\frac{g_s^2}{(2\pi)^3}\, 
\frac{(m+xM)^2\,(1-x)^3}
        {2\,[\bm{p}_{\sT}^2+L_s^2(\Lambda_s^2)]^4}  \; , 
\label{eq:f1+s} \\
f_1^{- q(s)} (x,\bm{p}_{\sT}^2) &\equiv \frac{1}{2}\Bigl(f_1^{q(s)}
(x,\bm{p}_{\sT}^2)  -g_{1L}^{q(s)} (x,\bm{p}_{\sT}^2)\Bigr)
=\frac{g_s^2}{(2\pi)^3}\, 
\frac{\bm{p}_{\sT}^2\,(1-x)^3}
       {2\,[\bm{p}_{\sT}^2+L_s^2(\Lambda_s^2)]^4}  \; , 
\label{eq:f1-s} \\
\nn \\
f_1^{+ q(a)} (x,\bm{p}_{\sT}^2) &\equiv \frac{1}{2}\Bigl(f_1^{q(a)}
(x,\bm{p}_{\sT}^2)  +g_{1L}^{q(s)} (x,\bm{p}_{\sT}^2)\Bigr)=\frac{g_a^2}{(2\pi)^3}\, 
\frac{\bm{p}_{\sT}^2\,(1+x^2)\,(1-x)}
        {2\,[\bm{p}_{\sT}^2 + L_a^2(\Lambda_a^2)]^4} \; , 
\label{eq:f1+a} \\
f_1^{- q(a)} (x,\bm{p}_{\sT}^2) &\equiv \frac{1}{2}\Bigl( f_1^{q(a)}
(x,\bm{p}_{\sT}^2)  -g_{1L}^{q(a)} (x,\bm{p}_{\sT}^2)\Bigr)=\frac{g_a^2}{(2\pi)^3}\, 
\frac{(m+xM)^2\,(1-x)^3}
        {2\,[\bm{p}_{\sT}^2 + L_a^2(\Lambda_a^2)]^4} \; . 
\label{eq:f1-a}
\end{align} 

Then, the TMDs can be written as (compare with Ref.~\cite{Bacchetta:2008af})
\begin{align} 
f_1^{q(s)}(x,\bm{p}_{\sT}^2) &= 
f_1^{+ q(s)} (x,\bm{p}_{\sT}^2)  +f_1^{- q(s)} (x,\bm{p}_{\sT}^2)  \; ,
\label{eq:f1sspect} \\
f_1^{q(a)} (x,\bm{p}_{\sT}^2) &= 
f_1^{+ q(a)} (x,\bm{p}_{\sT}^2)  +f_1^{- q(a)} (x,\bm{p}_{\sT}^2)   \; ,
\label{eq:f1aspect} \\
\nn \\
g_{1L}^{q(s)}(x,\bm{p}_{\sT}^2) &=
f_1^{+ q(s)} (x,\bm{p}_{\sT}^2)  -f_1^{- q(s)} (x,\bm{p}_{\sT}^2)  \; , 
\label{eq:g1Ls}  \\
g_{1L}^{q(a)}(x,\bm{p}_{\sT}^2) &= 
f_1^{+ q(a)} (x,\bm{p}_{\sT}^2)  -f_1^{- q(a)} (x,\bm{p}_{\sT}^2)   \; , 
\label{eq:g1La} \\ 
\nn \\
g_{1T}^{q(s)} (x,\bm{p}_{\sT}^2) &=
\frac{2 M}{m+xM}\, f_{1}^{+ q(s)} (x,\bm{p}_{\sT}^2) \; , 
\\
g_{1T}^{q(a)} (x,\bm{p}_{\sT}^2) &= 
\frac{x}{1-x}\frac{2 M}{m+xM}\, f_{1}^{- q(a)} (x,\bm{p}_{\sT}^2) \; , 
\label{eq:g1Tspect} \\ 
\nn \\
h_{1L}^{\perp\,q(s)}(x,\bm{p}_{\sT}^2) &= -g_{1T}^{q(s)}(x,\bm{p}_{\sT}^2) \; , 
\\
h_{1L}^{\perp\,q(a)}(x,\bm{p}_{\sT}^2) &= \frac{1}{x}\,g_{1T}^{q(a)}(x,\bm{p}_{\sT}^2) \; , 
\label{eq:h1Lperpspect} \\ 
\nn \\
h_{1T}^{q(s)}(x,\bm{p}_{\sT}^2) &= f_{1}^{q(s)}(x,\bm{p}_{\sT}^2) \; , 
\\
h_{1T}^{q(a)}(x,\bm{p}_{\sT}^2) &=  -\frac{2x}{1+x^2}\, f_{1}^{+ q(a)} (x,\bm{p}_{\sT}^2) \; , 
\label{eq:h1Tspect} \\ 
\nn \\
h_{1T}^{\perp\,q(s)}(x,\bm{p}_{\sT}^2)  &= -
\frac{2 M^2}{(m+xM)^2}\, f_{1}^{+ q(s)} (x,\bm{p}_{\sT}^2) \; , 
\\
h_{1T}^{\perp\,q(a)}(x,\bm{p}_{\sT}^2) &= 0 \; , 
\label{eq:h1Tperpspect} \\
\nn \\
h_1^{q(s)}(x,\bm{p}_{\sT}^2) &= f_{1}^{+ q(s)} (x,\bm{p}_{\sT}^2) \; , 
\\
h_1^{q(a)}(x,\bm{p}_{\sT}^2) &= h_{1T}^{q(a)}(x,\bm{p}_{\sT}^2) \; . 
\label{eq:h1spect}
\end{align} 

We stress that the previous relations among TMDs should be considered 
simply as a convenient way to write the results of our model, and should 
not be considered as general, as can be also deduced by the fact that they 
are all different in the scalar and axial-vector diquark case. For instance, 
the relation involving the pretzelosity $h_{1T}^{\perp}$, recently discussed by several 
authors~\cite{Avakian:2008dz,Meissner:2007rx,Efremov:2008mp,She:2009jq}, 
holds in the scalar diquark case, i.e. 
\begin{equation}
g_{1L}^s(x,\bm{p}_{\sT}) - h_1^s(x,\bm{p}_{\sT}) = 
\frac{\bm{p}_{\sT}^2}{2M^2}\, h_{1T}^{\perp\,s}(x,\bm{p}_{\sT}) \; ;
\label{eq:pretze}
\end{equation}
but it is not true for the axial-vector diquark, as it can be deduced
from Eqs.~(\ref{eq:g1La}), (\ref{eq:h1spect}), and (\ref{eq:h1Tperpspect}), by
noting that $h_{1T}^{\perp\,a}=0$ (see also Ref.~\cite{Bacchetta:2008af}). 
Therefore, such a relation cannot be totally general.

The ${\bm p}_{\sT}$-integrated results~\cite{Bacchetta:2008af} can be expressed in 
terms of the ${\bm p}_{\sT}$-integrated combinations~(\ref{eq:f1+s}-\ref{eq:f1-a}): 
\begin{align} 
f_1^{+q(s)} (x) &=\frac{g_s^2}{(2\pi)^2}\,
\frac{2(m+xM)^2\,(1-x)^3}{24\,L_s^6(\Lambda_s^2)}  \; , 
\label{eq:f1+sdip_} \\
f_1^{-q(s)} (x) &=\frac{g_s^2}{(2\pi)^2}\,
\frac{ (1-x)^3}{24\,L_s^4(\Lambda_s^2)}  \; , 
\label{eq:f1-sdip_} \\
\nn \\
f_1^{+q(a)} (x) &=\frac{g_a^2}{(2\pi)^2}\,
\frac{(1+x^2)\, (1-x)}{24\,L_a^4(\Lambda_a^2)}  \; , 
\label{eq:f1+adip_} \\
f_1^{-q(a)} (x) &=\frac{g_a^2}{(2\pi)^2}\,
\frac{2(m+xM)^2\,(1-x)^3}{24\,L_a^6(\Lambda_a^2)}  \; . 
\label{eq:f1-adip_} \\ 
\end{align} 
The final result is 
\begin{align} 
f_1^{q(s)} (x) &= f_1^{+q(s)} (x) + f_1^{-q(s)} (x) \; , 
\label{eq:f1sdip_int} \\
f_1^{q(a)} (x) &= f_1^{+q(a)} (x) + f_1^{-q(a)} (x) \; , 
\label{eq:f1adip_int} \\ 
\nn \\
g_1^{q(s)}(x) &= f_1^{+q(s)} (x) - f_1^{-q(s)} (x) \; , 
\label{eq:g1sdip_int}  \\
g_1^{q(a)}(x) &= f_1^{+q(a)} (x) - f_1^{-q(a)} (x)  \; , 
\label{eq:g1adip_int}  \\ 
\nn \\
h_1^{q(s)}(x) &=  f_1^{+q(s)} (x) \; , 
\label{eq:h1sdip_int} \\
h_1^{q(a)}(x) &= - \frac{2x}{1+x^2}\,  f_1^{+q(a)} (x) \; .
\label{eq:h1adip_int} 
\end{align} 

The weighted SSA of Eqs.~(\ref{eq:Altcosph-S}), (\ref{eq:Aulsin2ph}), (\ref{eq:Autsin3ph-S}), 
and (\ref{eq:dyAutsin3ph-S}), require the knowledge also of the
following $\bm{p}_{\sT}$-moments: 
\begin{align}
g_{1T}^{q(s)\, (1)}(x) &=\frac{g_s^2}{(2\pi)^2}\, 
\frac{(m+xM)\,(1-x)^3}{24\, M\, L_s^4(\Lambda_s^2)} \; , \label{eq:g1Ts1} \\
g_{1T}^{q(a)\, (1)}(x) &=\frac{g_a^2}{(2\pi)^2}\, 
\frac{(m+xM)\, x\, (1-x)^2}{24\, M\, L_a^4(\Lambda_a^2)} \; , \label{eq:g1Ta1} 
\\
\nn \\
h_{1L}^{\perp q(s)\, (1)}(x) &= - g_{1T}^{q(s)\, (1)}(x) \; , 
\label{eq:h1Ls1} \\
h_{1L}^{\perp q(a)\, (1)}(x) &= \frac{1}{x}\, g_{1T}^{q(a)\, (1)}(x) \; , 
\label{eq:h1La1} \\ 
\nn \\
h_{1T}^{\perp q(s)\, (2)}(x) &= - \frac{g_s^2}{(2\pi)^2}\, 
\frac{(1-x)^3}{24\, M^2\, L_s^2(\Lambda_s^2)} \; , \label{eq:h1Ts2} \\
h_{1T}^{\perp q(a)\, (2)}(x) &= 0 \; . \label{eq:h1Ta2}
\end{align}

For completeness, we also list the $\bm{p}_{\sT}$-moments of the 
combinations~(\ref{eq:f1+s}-\ref{eq:f1-a}), which are useful to calculate the mean squared 
transverse momentum associated to each of the above TMDs: 
\begin{align}
f_1^{+q(s)\, (1)} (x) &=\frac{g_s^2}{(2\pi)^2}\,
\frac{(m+xM)^2\,(1-x)^3}{96\,M^2\,L_s^4(\Lambda_s^2)}  \; , 
\label{eq:f1+sdip_(1)} \\
f_1^{+q(a)\, (1)} (x) &=\frac{g_a^2}{(2\pi)^2}\,
\frac{(m+xM)^2\,(1-x)^3}{96\,M^2\,L_a^4(\Lambda_a^2)}  \; , 
\label{eq:f1+adip_(1)} \\ 
\nn \\
f_1^{-q(s)\, (1)} (x) &=\frac{g_s^2}{(2\pi)^2}\,
\frac{2 \,(1-x)^3}{96\,M^2\,L_s^2(\Lambda_s^2)}  \; , 
\label{eq:f1-sdip_(1)} \\
f_1^{-q(a)\, (1)} (x) &=\frac{g_a^2}{(2\pi)^2}\,
\frac{2\,(1+x^2)\,(1-x)}{96\,M^2\,L_a^2(\Lambda_a^2)}  \; . 
\label{eq:f1-adip_(1)}
\end{align}

\begin{figure}[h]
\begin{center}
\includegraphics[width=6cm]{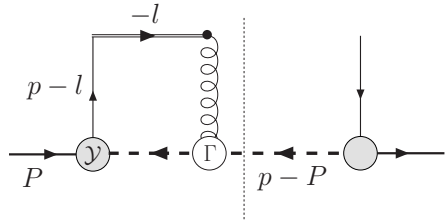}
\end{center}
\caption{Interference between the one-gluon exchange diagram in eikonal
  approximation and the tree level diagram in the spectator model. The
  Hermitean conjugate diagram is not shown.}
\label{fig:eikonal}
\end{figure}


\subsection{T-odd}
\label{sec:todd}

There are only two T-odd structures that can be extracted at leading twist from
the quark-quark correlator of Eq.~(\ref{eq:Phi-tree}), namely 
\begin{align} 
\frac{\varepsilon_{\sT}^{ij}p_{\sT i}S_{\sT j}}{M}\,
f_{1T}^{\perp}(x,\bm{p}_{\sT}^2) &= -\frac{1}{4}\, \mathrm{Tr} \left[ \left( 
\Phi (x,\bm{p}_{\sT},S) -\Phi (x,\bm{p}_{\sT},-S)\right) \, \g^+ \right] + 
\mathrm{h.c.} \; , \label{eq:Sivers} \\
\frac{\varepsilon_{\sT}^{ij}p_{\sT j}}{M}\,
h_1^{\perp}(x,\bm{p}_{\sT}^2) &= \frac{1}{4}\, \mathrm{Tr} \left[ \left( 
\Phi (x,\bm{p}_{\sT},S) +\Phi (x,\bm{p}_{\sT},-S)\right) \,
i\sigma^{i+}\,\g_5 \right] + \mathrm{h.c.} \; , \label{eq:Boer-Mulders} 
\end{align}
which are usually referred to as the Sivers~\cite{Sivers:1990cc} and 
Boer-Mulders~\cite{Boer:1997nt} functions, respectively. 

Using the diagram rules for $\mathcal{M}^{(0)}$ in Fig.~\ref{fig:sidis}, from
Eq.~(\ref{eq:Phi-tree-spect}) we find that these expressions vanish at tree 
level, because there is no interference between two competing channels 
producing the complex amplitude whose imaginary part gives the T-odd 
contribution. We can generate such structures by considering the interference 
between the tree-level scattering amplitude $\mathcal{M}^{(0)}$ and the 
single-gluon-exchange scattering amplitude $\mathcal{M}^{(1)}$ in eikonal 
approximation, as shown in Fig.~\ref{fig:eikonal} (the Hermitean conjugate 
partner must also be considered). This corresponds to the 
one-gluon-exchange approximation of the gauge link operator of 
Eq.~(\ref{eq:link})~\cite{Boer:2003cm}. 

The Sivers and Boer-Mulders functions can then be rewritten as
\begin{align} 
\frac{\varepsilon_{\sT}^{ij}p_{\sT i}S_{\sT j}}{M}\,
f_{1T}^{\perp}(x,\bm{p}_{\sT}^2) &= -\frac{1}{4}\,\frac{1}{(2\pi)^3}\,
\frac{1}{2(1-x)P^+}\,\mathrm{Tr}\, \left[ \left(
\mathcal{M}^{(1)}(S) \overline{\mathcal{M}}^{(0)}(S) - \mathcal{M}^{(1)}(-S) 
\overline{\mathcal{M}}^{(0)}(-S) \right)\, \g^+ \right] + \mathrm{h.c.} \; , 
\label{eq:Siversspect} \\
\frac{\varepsilon_{\sT}^{ij}p_{\sT j}}{M}\,
h_1^{\perp}(x,\bm{p}_{\sT}^2) &= \frac{1}{4}\,\frac{1}{(2\pi)^3}\,
\frac{1}{2(1-x)P^+}\,\mathrm{Tr}\, \left[ \left(
\mathcal{M}^{(1)}(S) \overline{\mathcal{M}}^{(0)}(S) + \mathcal{M}^{(1)}(-S) 
\overline{\mathcal{M}}^{(0)}(-S) \right)\, i\sigma^{i+}\,\g_5 \right] + 
\mathrm{h.c.} \; . 
\label{eq:BoerMuldersspect}
\end{align} 
Following the lines of Ref.~\cite{Bacchetta:2008af}, we come to the final 
form with scalar and axial vector diquarks: 
\begin{align}
f_{1T}^{\perp\,q(s)}(x,\bm{p}_{\sT}^2) &= -g_s^2 C_F \alpha_s\,
\frac{M\,\pi}{(2\pi)^4}\,\frac{(1-x)^3\,(m+xM)}
     {L_s^2(\Lambda_s^2)\,[\bm{p}_{\sT}^2+L_s^2(\Lambda_s^2)]^3} \; , 
\label{eq:Siversspect3s} 
\\
f_{1T}^{\perp\,q(a)}(x,\bm{p}_{\sT}^2) &= g_a^2 C_F \alpha_s\,
\frac{M\,\pi}{(2\pi)^4}\,\frac{(1-x)^2\,x\,(m+xM)}
     {L_a^2(\Lambda_a^2)\,[\bm{p}_{\sT}^2+L_a^2(\Lambda_a^2)]^3} \; , 
\label{eq:Siversspect3a} 
\\ \nn \\
h_1^{\perp\,q(s)}(x,\bm{p}_{\sT}^2) &= f_{1T}^{\perp\,q(s)}(x,\bm{p}_{\sT}^2)
\label{eq:BoerMuldersspect3s} \; , 
\\
h_1^{\perp\,q(a)}(x,\bm{p}_{\sT}^2) &=-\frac{1}{x}\,
f_{1T}^{\perp\,q(a)}(x,\bm{p}_{\sT}^2) \; , 
\label{eq:BoerMuldersspect3a}
\end{align} 
with $C_F = 4/3$. 

The high$-\bm{p}_{\sT}$ tail of above expressions could break the positivity
bounds~\cite{Bacchetta:1999kz}. However, the spectator model is supposed to be 
valid for $\bm{p}_{\sT}^2 \sim M^2$ and for reasonable choices of the 
parameters no problems with positivity occurr in this region (for more details, 
see Ref.~\cite{Bacchetta:2008af}). 

Finally, in the weighted SSA we need the first $\bm{p}_{\sT}$ moments of such 
functions, which read ~\cite{Bacchetta:2008af}
\begin{align}
f_{1T}^{\perp\,q(s)\,(1)}(x) 
&= -\frac{g_s^2}{8}\,\frac{\pi\,C_F \alpha_s}{(2\pi )^3\,M}\,
   \frac{(m+xM)\,(1-x)^3}{[L_s^2(\Lambda_s^2)]^2} \; , 
\label{eq:Sivers1s} \\
f_{1T}^{\perp\,q(a)\,(1)}(x) 
&= \frac{g_a^2}{8}\,\frac{\pi\,C_F \alpha_s}{(2\pi )^3\,M}\,
   \frac{x\,(m+xM)\,(1-x)^2}{[L_a^2(\Lambda_a^2)]^2} \; , 
\label{eq:Sivers1a} \\ 
\nn \\
h_1^{\perp\,q(s)\,(1)}(x) &= f_{1T}^{\perp\,q(s)\,(1)}(x) 
\label{eq:BoerMulders1s} \; , 
\\
h_1^{\perp\,q(a)\,(1)}(x) &= -\frac{1}{x}\,f_{1T}^{\perp\,q(a)\,(1)}(x) \; .
\label{eq:BoerMulders1a} 
\end{align}


\section{Results and predictions}
\label{sec:out}

In the following, we present the results for the weighted SSA discussed in 
Sec.~\ref{sec:ssa}. We recall that the model parameters were fixed in 
Ref.~\cite{Bacchetta:2008af} by fitting the ZEUS parametrization for $f_1^u$ 
and $f_1^d$ at the lowest available scale $Q_0^2= 0.3$ 
GeV$^2$~\cite{Chekanov:2002pv}, and the GRSV2000 parametrization at LO for 
$g_1^u$ and $g_1^d$ at $Q_0^2 = 0.26$ GeV$^2$~\cite{Gluck:2000dy}. The 
connection with the model TMDs is realized by 
\begin{align}
f_1^u &= c_{s}^2\, f_1^{u(s)} + c_{a}^2\, f_1^{u(a)} \nn \\
f_1^d &= c_a^{\prime 2}\, f_1^{d(a')}  \; ,
\label{eq:qX}
\end{align} 
with the coefficients $c_X^{(\prime )}$ given in Tab.I of 
Ref.~\cite{Bacchetta:2008af}. They give the relative probability weight of each 
quark-diquark configuration in spin and isospin space: scalar-isoscalar diquark 
$s$ with quark $u$ ($c_s$), axial vector-isovector diquark $a$ with isospin projection 0 
and quark $u$ ($c_a$), axial vector-isovector diquark $a'$ with isospin projection 1 
and quark $d$ ($c_a'$). 

The model (hadronic) scale is assumed to be $Q_0^2=0.3$ GeV$^2$. At this scale, 
the contribution of sea over valence quarks is approximately within 10-20\% for 
$x\gtrsim 0.1$; therefore, we have involved only valence contributions in the fitting 
procedure, keeping in mind that there will be discrepancies at $x<0.1$. Viceversa, 
for the $D_1$ fragmentation function we employ the parametrization of 
Ref.~\cite{deFlorian:2007aj} where both favoured and unfavoured channels are included, 
the latter being particularly important even at low $Q^2$. For the Collins 
function $H_1^{\perp}$ in Eq.~(\ref{eq:Autsinph+S}), we adopt the analytic expression 
of Ref.~\cite{Bacchetta:2007wc}, which was obtained in a similar spectator approach
as our model TMDs. However, in this case the unfavoured contributions are 
simply assumed to be equal and opposite to the favoured ones. 

Evolution with running $Q^2$ is implemented by suitably extending the HOPPET 
code of Ref.~\cite{Salam:2008qg} to include also chiral-odd partonic densities. 
The modification is here considered at LO only, and the evolution of all 
partonic densities will be considered at LO as well in the following, because 
the factorized expressions of the SSA in 
Eqs.~(\ref{eq:Autsinph+S})--(\ref{eq:Altcosph-S}) 
and~(\ref{eq:dyAutsinph-S}),(\ref{eq:dyAutsinph+S}) are valid at LO only.

In the following subsection, we reconsider evolution properties of our T-odd
TMDs. In the next two subsections, we will present our results for weighted SSA 
in the SIDIS and in the (polarized) Drell--Yan processes, respectively.


\subsection{Results for model T-odd TMDs}
\label{sec:tmd-out}

As already anticipated in Sec.~\ref{sec:sidis}, evolution equations are not 
available for all partonic densities in the equations for the considered SSA. At LO, 
the DGLAP equations for $f_{1T}^{\perp\, (1)} (x), \, g_{1T}^{(1)} (x), \, h_1^{\perp\, (1)}(x),$ 
are assumed to be the same as for $f_1(x), \, g_1(x), \, h_1(x),$ respectively. 

Moreover, the first $\bm{p}_{\sT}$ moments of these T-odd TMDs depend linearly 
on $\alpha_s$ [see Eqs.~(\ref{eq:Sivers1s})-(\ref{eq:BoerMulders1a})]. 
Therefore, a consistent treatment of evolution effects requires to deduce from the LO 
renormalization group equations the value of $\alpha_s$ at the scale of 
the hadronic model, $Q_0^2$, while $\alpha_s(Q_0^2)$ is
often considered as a free parameter. 

In this sense, in the following we update the results for the model T-odd TMDs 
with respect to our previous paper~\cite{Bacchetta:2008af} by replacing the {\it ad-hoc} 
parameter $\alpha_s(Q_0^2)=0.3$ with the value of $\alpha_s(Q_0^2)=0.697$ 
deduced from the LO running of the strong coupling starting from 
a value of $\alpha(M_Z^2)=0.125$ (see discussion in Ref.~\cite{Gluck:1998xa}) 
and using the standard choices of the 
HOPPET evolution program~\cite{Salam:2008qg}.

\begin{figure}[h]
\begin{center}
\includegraphics[width=7cm]{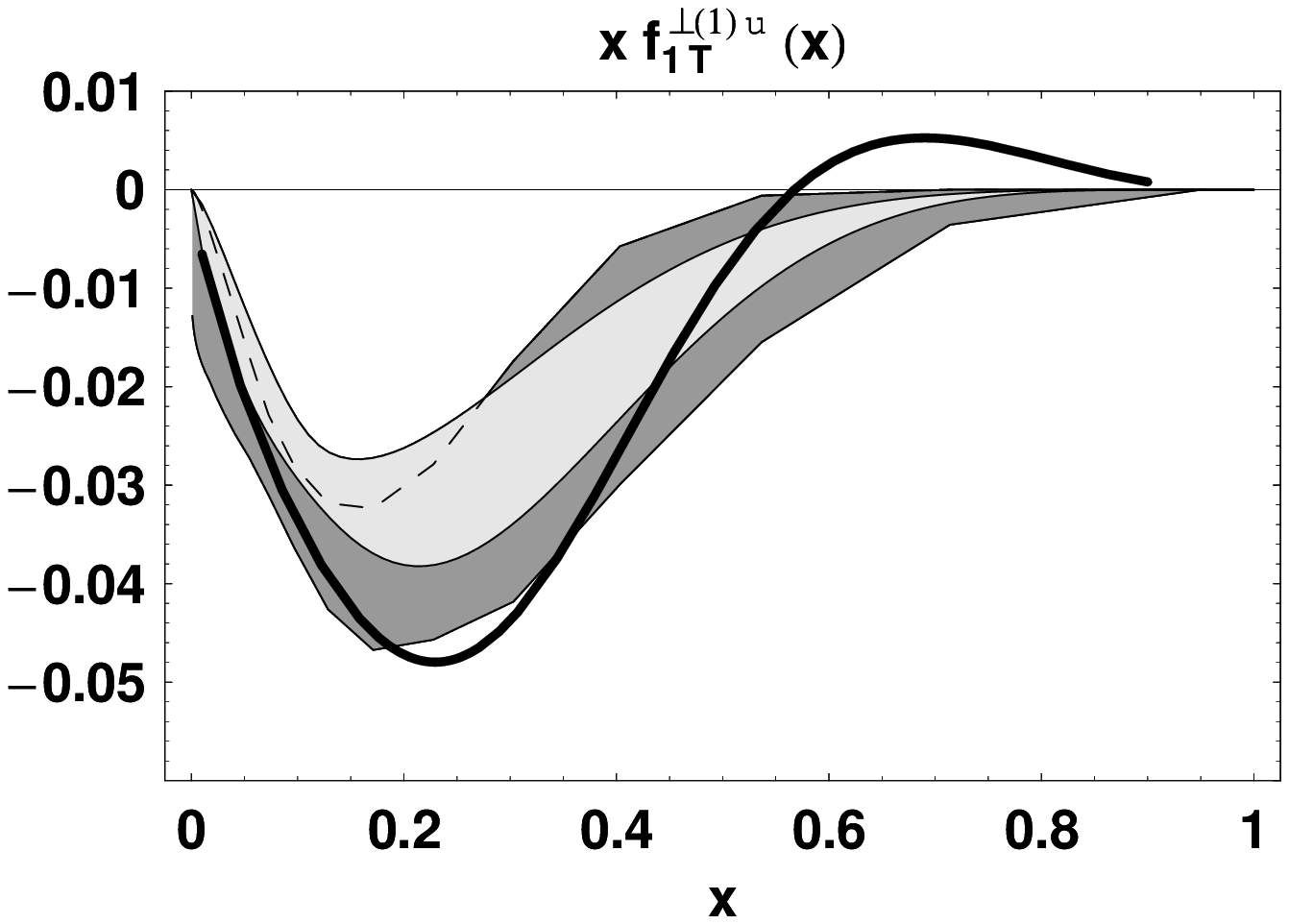} \hspace{0.2cm} 
\includegraphics[width=7cm]{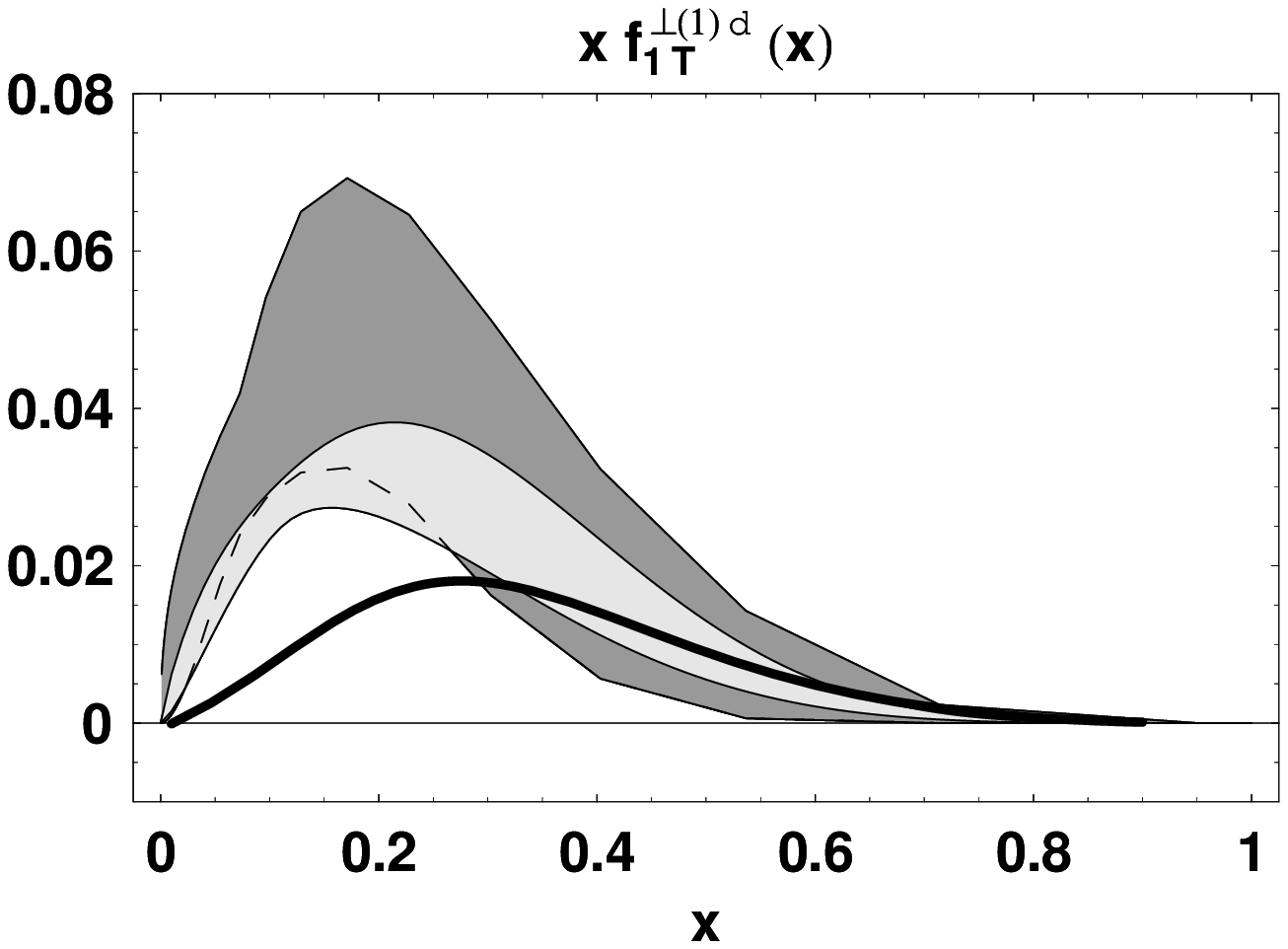} 
\end{center}
\caption{The first ${\bm p}_{\sT}$ moment $x f_{1T}^{\perp\, (1)}(x)$ of the 
Sivers function; left (right) panel for up (down) quark. Darker shaded area: 
statistical uncertainty band of the parametrizations from 
Ref.~\protect{\cite{Anselmino:2008sg}}, lighter shaded area from 
Ref.~\protect{\cite{Collins:2005wb}}. Solid line for the results of the 
spectator diquark model at the parametrization scale $Q^2=2.5$ GeV$^2$.}
\label{fig:sivers1}
\end{figure}

In Fig.~\ref{fig:sivers1}, the first ${\bm p}_{\sT}$ moment 
$x f_{1T}^{\perp\, (1)}(x)$ of the Sivers function is displayed for the up and
down valence quarks in the left and right panels, respectively. The darker 
shaded area represents the uncertainty due to the statistical error in the 
parametrization of Ref.~\cite{Anselmino:2008sg}, while the lighter shaded area 
corresponds to the same in Ref.~\cite{Collins:2005wb}. Both parametrizations
are deduced by fitting data for the Sivers effect measured by the 
HERMES~\cite{Diefenthaler:2007rj} and COMPASS~\cite{Martin:2007au,:2008dn} 
collaborations, hence at an average scale $Q^2=2.5$ GeV$^2$. The solid line is 
given by combining the model results of Eqs.~(\ref{eq:Sivers1s}) 
and~(\ref{eq:Sivers1a}) according to Eq.~(\ref{eq:qX}), and by further evolving 
them at LO from the model scale $Q_0^2=0.3$ GeV$^2$ to the above mentioned 
parametrization scale. 

First of all, we observe the agreement between the signs of the various flavor 
components, which also agree with the findings from calculations on the 
lattice~\cite{Gockeler:2006zu}. The agreement between model and parametrizations
is very good for the up quark, even if the maximum is slightly shifted towards 
higher $x$. The size of the function is instead too small for the down 
quark, and its shape shifted such that the maximum occurs at a higher value of 
$x\approx 0.3$.

\begin{figure}[h]
\begin{center}
\includegraphics[width=7cm]{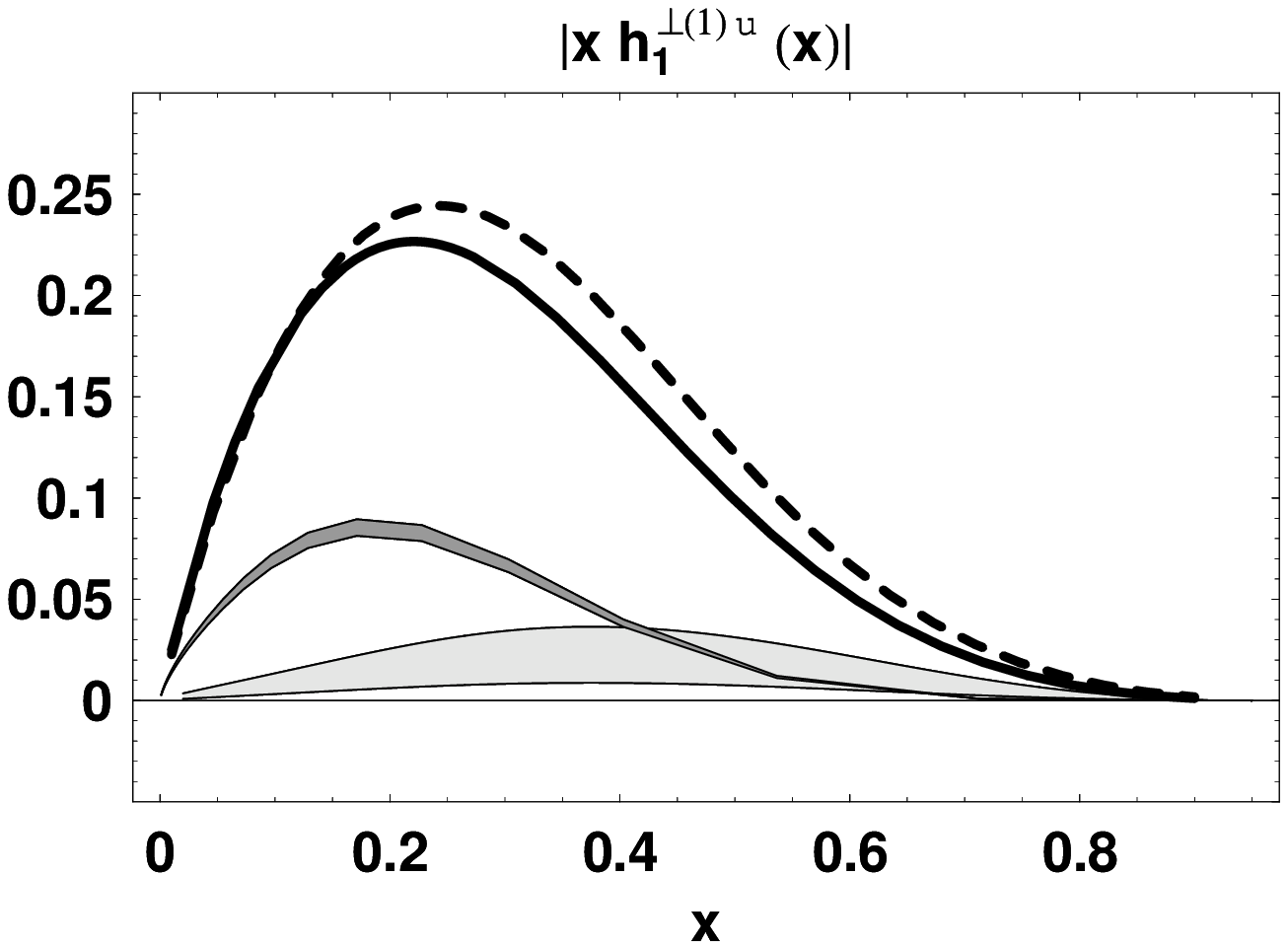} \hspace{0.2cm} 
\includegraphics[width=7cm]{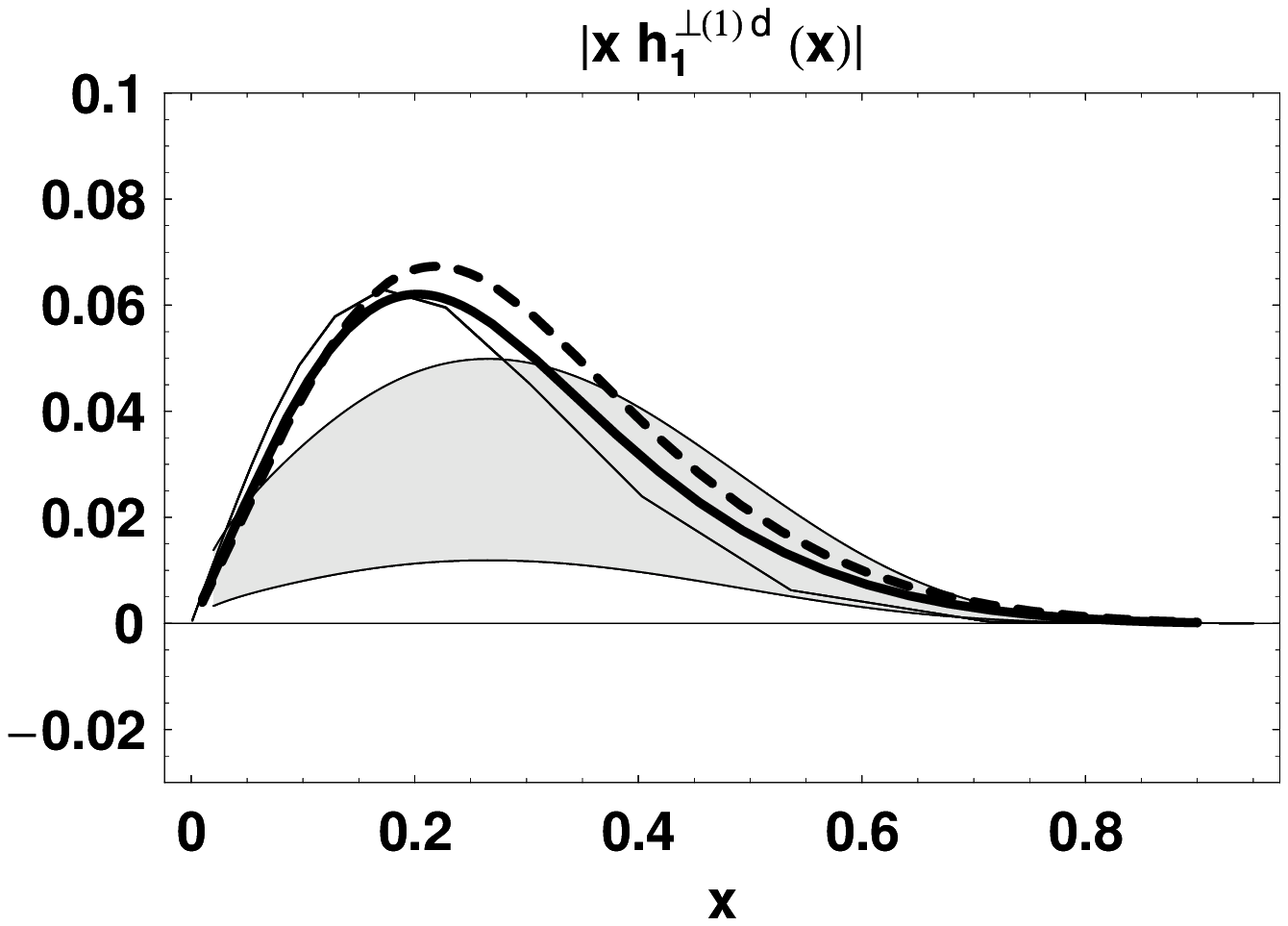} 
\end{center}
\caption{The modulus of the first ${\bm p}_{\sT}$ moment 
$|x h_1^{\perp\, (1)}(x)|$ of the Boer--Mulders function; left (right) panel for 
up (down) quark. Darker shaded area: statistical uncertainty band of the parametrization from 
Ref.~\protect{\cite{Barone:2009hw}}, extracted from SIDIS data at $Q^2=2.5$ 
GeV$^2$. Lighter shaded area: error assigned to parametrization from 
Ref.~\protect{\cite{Lu:2009ip}}, extracted from Drell--Yan data at $Q^2=1$ 
GeV$^2$. Solid and dashed lines for the results of the spectator diquark model 
evolved at $Q^2=2.5$ and $1$ GeV$^2$, respectively. We remark that the results 
of the spectator diquark model are negative for both up and down quarks (see 
Ref.~\protect{\cite{Bacchetta:2008af}}); here, we plot their modulus in order to conveniently 
compare with the published parametrizations.}
\label{fig:bm1}
\end{figure}

In Fig.~\ref{fig:bm1}, the modulus of the first ${\bm p}_{\sT}$ moment 
$|x h_1^{\perp\, (1)}(x)|$ of the Boer-Mulders function is displayed for the up 
and down valence quarks in the left and right panels, respectively. The darker 
shaded area represents the uncertainty due to the statistical error in the 
parametrization of Ref.~\cite{Barone:2009hw}, deduced by fitting the 
$\cos 2\phi$ asymmetry in the unpolarized SIDIS data of 
Ref.~\cite{Kafer:2008ud,Bressan:2009eu,Giordano:2009hi} at $Q^2=2.5$ GeV$^2$ 
[see Eq.~(\ref{eq:Auucos2ph})]. For the down quark, the uncertainty in the
fitting parameter is so small that the area reduces to a thin line. The lighter 
shaded area corresponds to the similar analysis from Ref.~\cite{Lu:2009ip} but 
for the unpolarized Drell--Yan data of Ref.~\cite{Zhu:2008sj} at $Q^2=1$ GeV$^2$ 
[see Eq.~(\ref{eq:dyAuucos2ph})]. The solid line is given by combining now the 
model results of Eqs.~(\ref{eq:BoerMulders1s}) and~(\ref{eq:BoerMulders1a}) 
according to Eq.~(\ref{eq:qX}), and by further evolving them at LO from the 
model scale $Q_0^2=0.3$ GeV$^2$ to $Q^2=2.5$ GeV$^2$. The dashed line displays 
what we obtain when we perform the LO evolution up to $Q^2=1$ GeV$^2$. Hence, 
the solid line must be compared with the darker shaded area, while the dashed 
line with the lighter shaded area. 

We remark that the $\cos 2\phi$ asymmetry in Drell--Yan involves two Boer--Mulders 
functions and the fitting procedure cannot fix the sign of the function depending 
on the flavor involved. The $\cos 2\phi$ asymmetry in SIDIS involves 
$h_1^{\perp}$ in combination with the unknown Collins function, whose sign in turn depends 
on the sign of the transversity $h_1$. The extraction of Ref.~\cite{Barone:2009hw} 
assumes the favoured Collins function to be positive and obtains negative 
Boer--Mulders functions for both $u$ and $d$ quarks, in agreement with our 
model~\cite{Bacchetta:2008af}, with lattice calculations~\cite{Gockeler:2006zu},  
and also with other models~\cite{Gamberg:2007wm,Yuan:2003wk}. The size of 
$|x h_1^{\perp\, (1)u}(x)|$ is too high in Fig.~\ref{fig:bm1}, while the
result for the down quark seems in better agreement with the SIDIS 
parametrization than with the Drell--Yan one. However, the comparison should be 
considered with some care because our model results contain only the pure
contribution from valence quarks and the data are contaminated by perturbative
QCD contributions and higher twists (e.g., the Cahn effect in SIDIS).


\subsection{The SSA in SIDIS}
\label{sec:sidis-out}

In the following, all displayed experimental data for weighted SSA in SIDIS 
were collected by the HERMES collaboration at the cm energy squared $s=56.2$ 
GeV$^2$ and the average scale $\langle Q^2\rangle = 2.5$ GeV$^2$ (but evolution 
effects were included for each $Q^2(x)$ in the $x$ distribution). Applied 
experimental cuts are: 
\begin{align}
& 0.023<x<0.4 \; , &0.2<z<0.8 \; , \hspace{5.5cm} \nn \\
& y_{\text{min}}(x)<y<0.85 \; , &y_{\text{min}}(x)=\text{max} \left[ 0.1,\, 
\frac{1}{x (s-M^2)}, \, \frac{4-M^2}{(1-x)\, (s-M^2)} \right] \; . 
\label{eq:sidiskin}
\end{align}
Particular care must be taken when considering the asymmetry as a function
of $z$, namely when separately integrating its numerator and denominator as 
functions of $x$ and $y$. 

\begin{figure}[h]
\begin{center}
\includegraphics[width=6cm]{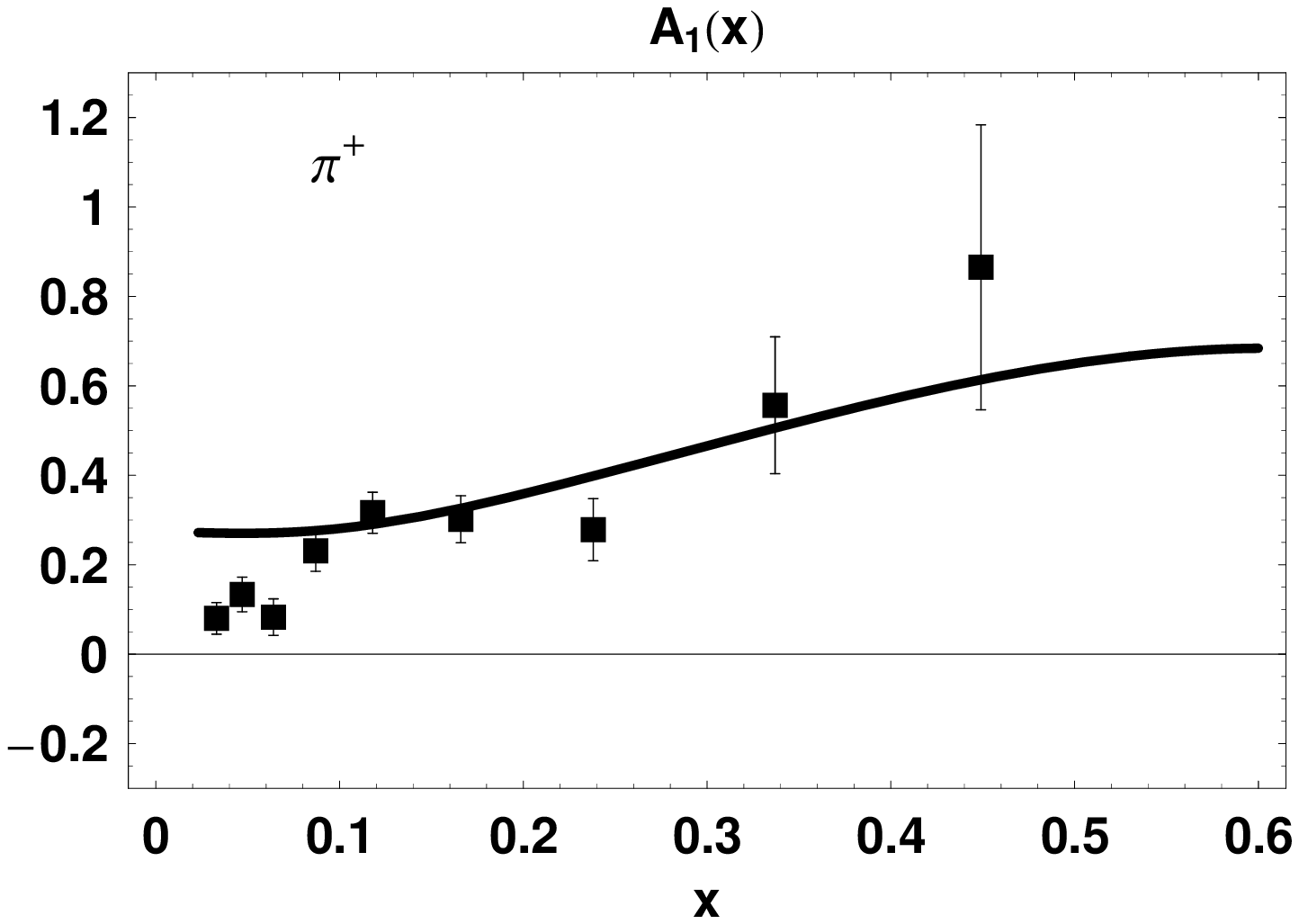} \hspace{0.2cm} 
\includegraphics[width=6cm]{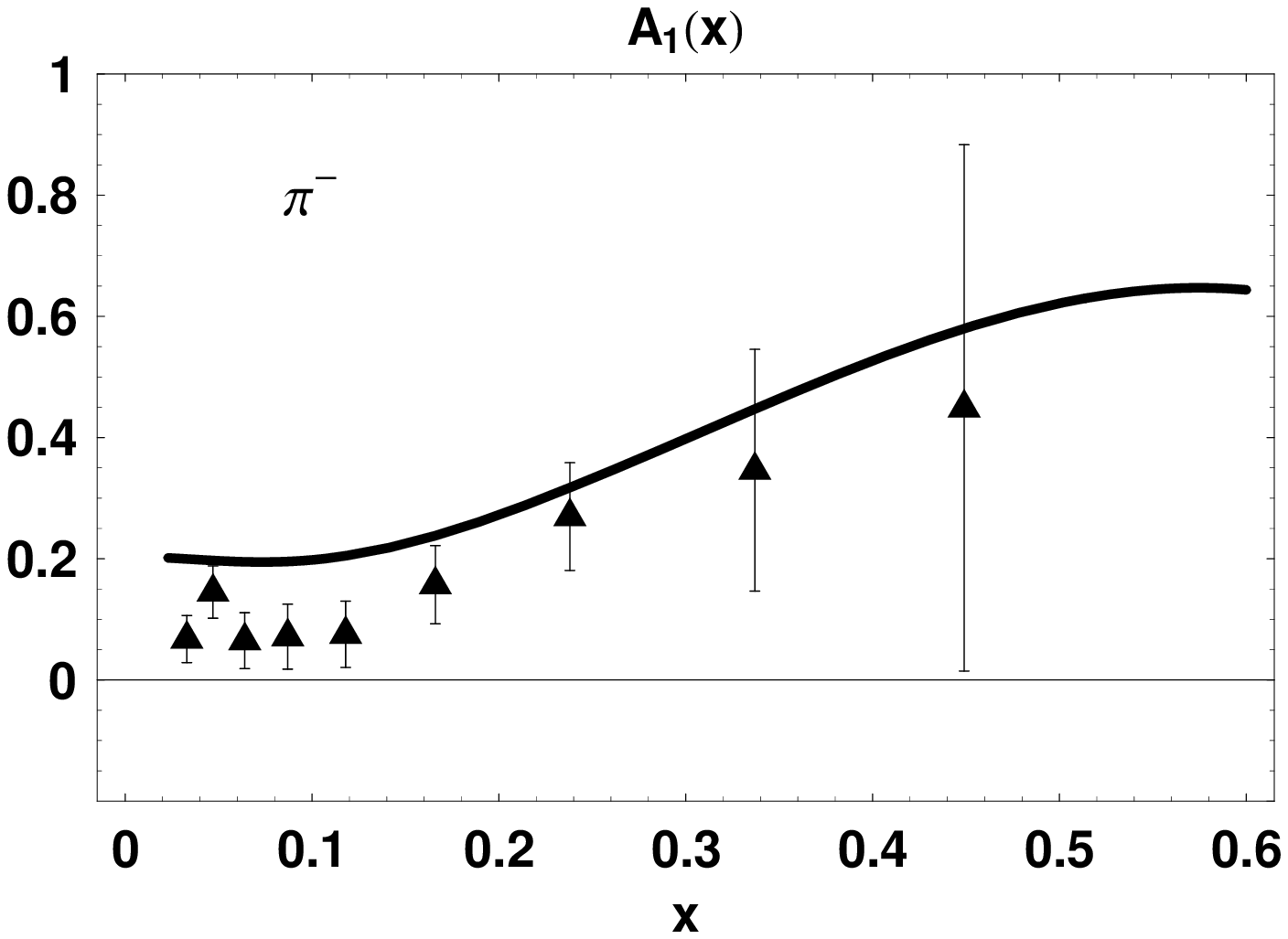} 
\end{center}
\caption{The double-spin asymmetries $A_1(x)$ from Eq.~(\protect{\ref{eq:all}}) (see text) 
in the SIDIS kinematics of HERMES with emission of $\pi^+$ (left) and $\pi^-$ (right). 
Experimental data from Ref.~\cite{Airapetian:2004zf}. Solid line represents the result of the spectator model.}
\label{fig:all}
\end{figure}

In Fig.~\ref{fig:all}, we show the $x$ dependence of the virtual photon asymmetry 
$A_1$ from Eq.~(\ref{eq:all}). Experimental data for both $\pi^+$ (left) and 
$\pi^-$ emission (right) are taken from Ref.~\cite{Airapetian:2004zf} in the
above mentioned SIDIS kinematics of HERMES. 

In all cases, the solid line is the result of our spectator model when employing
the fragmentation function $D_1$ of Ref.~\cite{deFlorian:2007aj}, including LO
evolution of all partonic densities to the experimental scale. The agreement 
with data is satisfactory. The deviations from data at low $x$ are driven by contributions of 
nonvalence partons (sea quarks), which are not included in the present version 
of the model (see the comment at the beginning of Sec.~\ref{sec:out}).

\begin{figure}[h]
\begin{center}
\includegraphics[width=5.5cm]{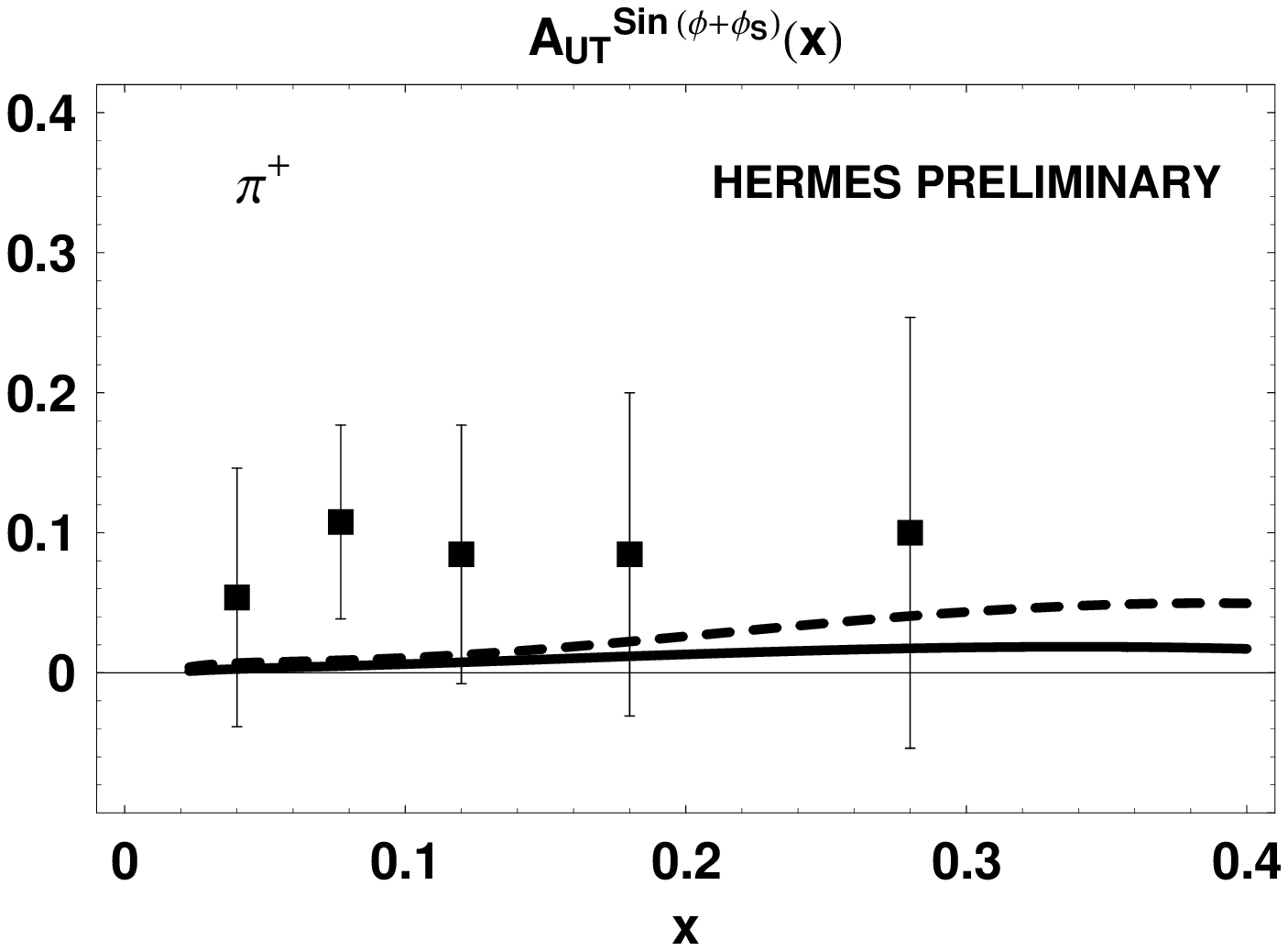} \hspace{0.2cm} 
\includegraphics[width=5.5cm]{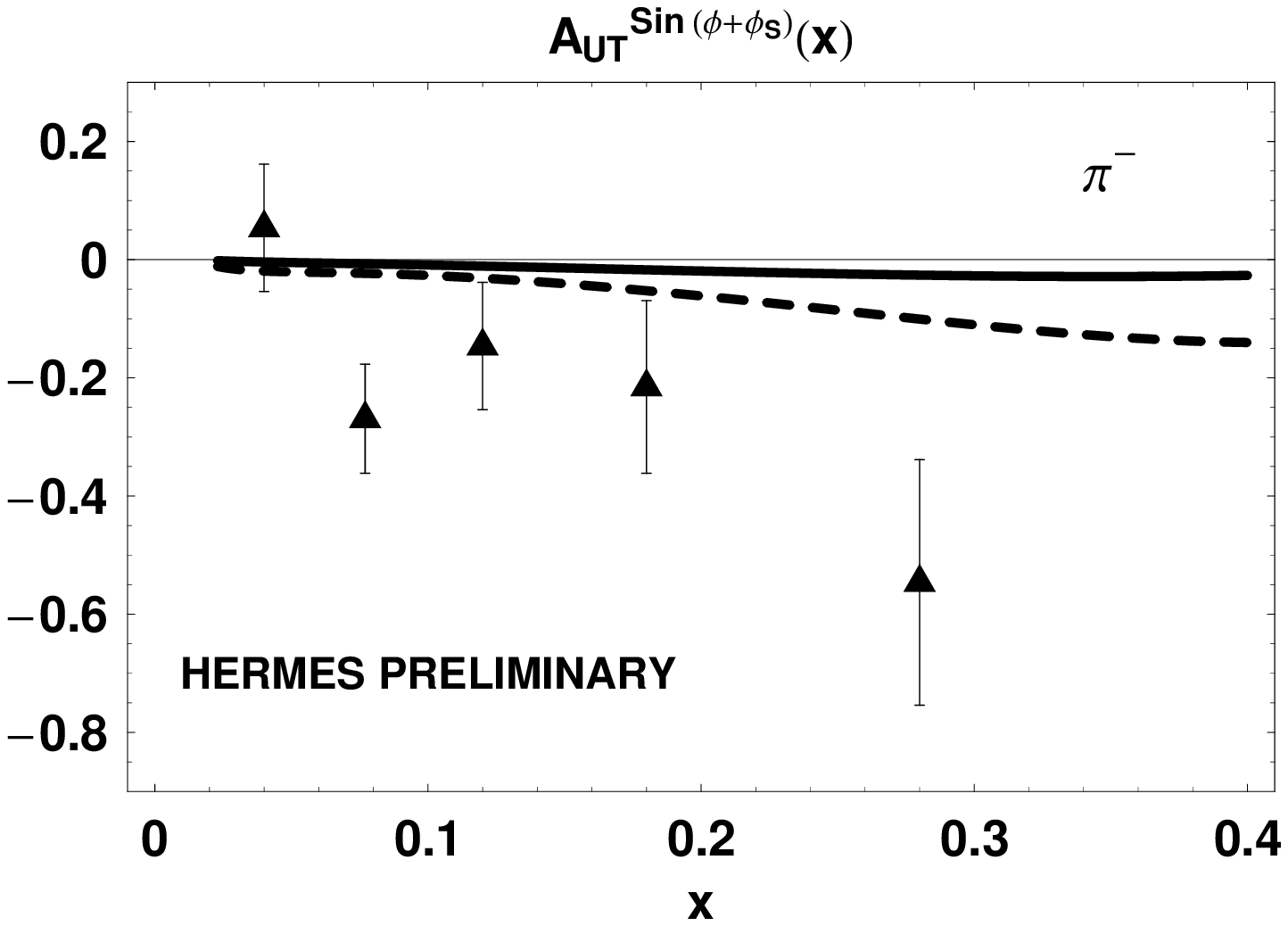} \\[0.2cm]
\includegraphics[width=5.5cm]{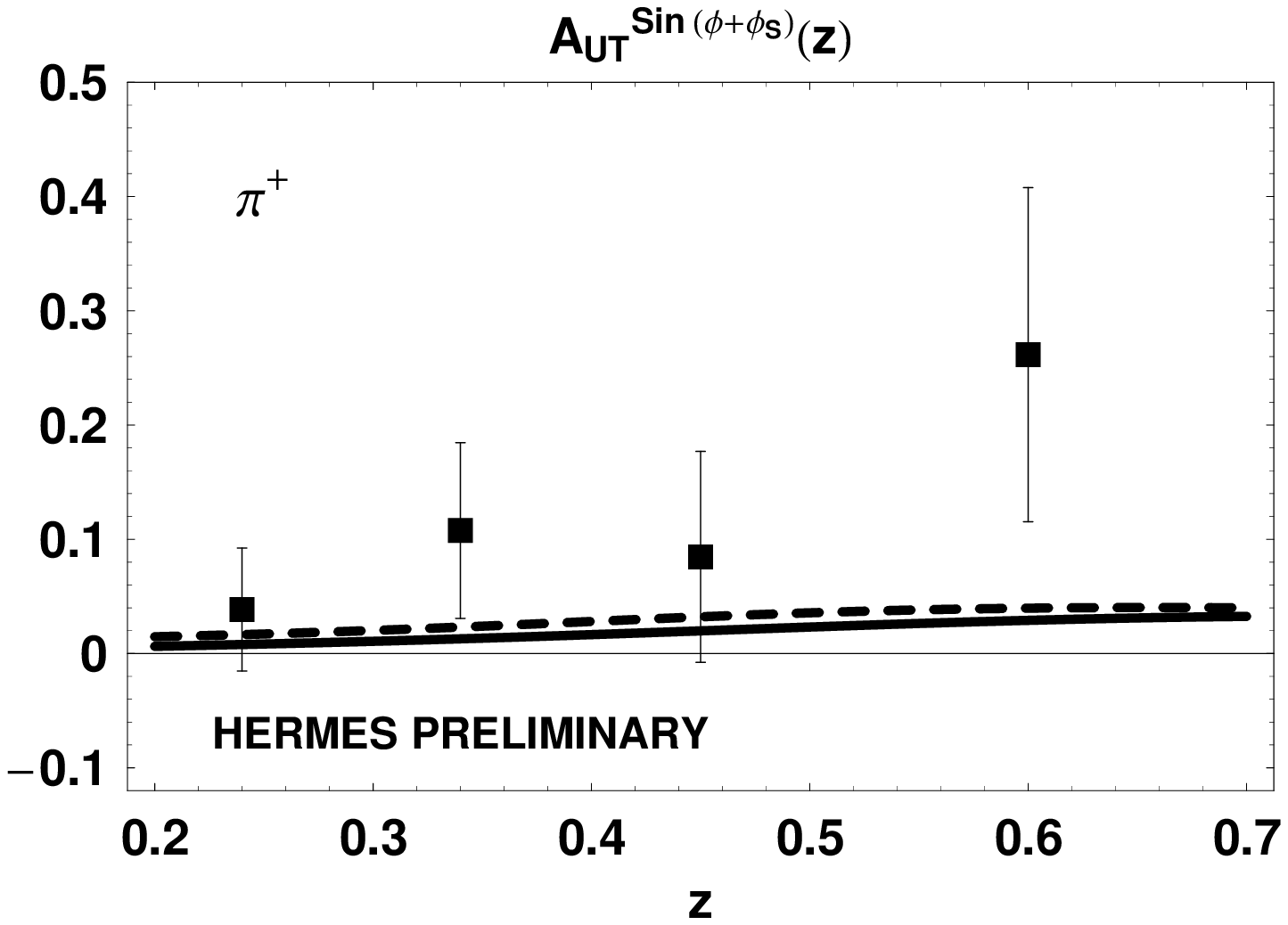} \hspace{0.2cm} 
\includegraphics[width=5.5cm]{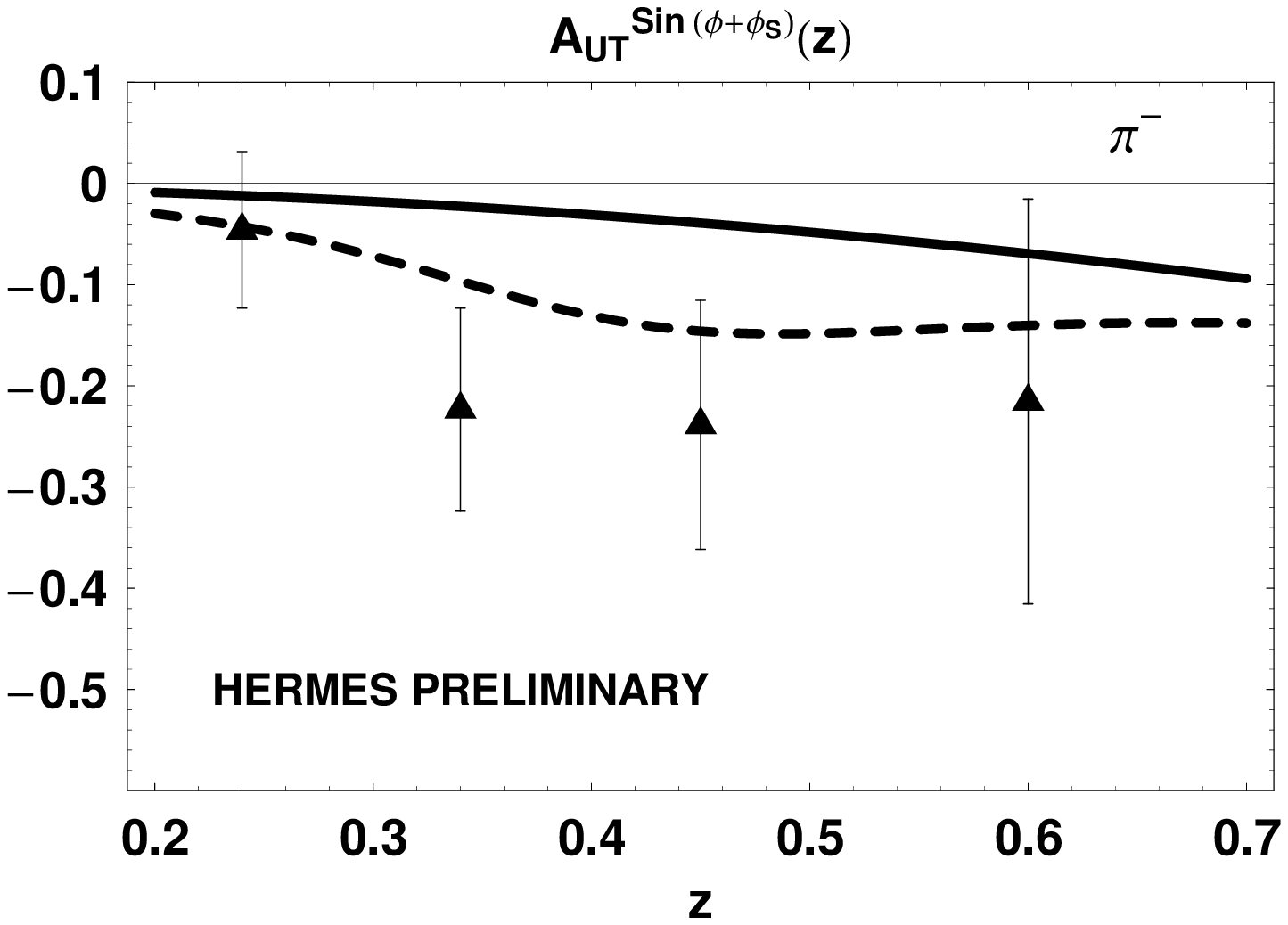}
\end{center}
\caption{The weighted single-spin asymmetry 
$A_{UT}^{Q_T\,\sin (\phi_h+\phi_S)}$ of Eq.~(\protect{\ref{eq:Autsinph+S}}) in 
the SIDIS kinematics of HERMES (Collins effect) with emission of $\pi^+$ (left) 
and $\pi^-$ (right), as a function of $x$ (above) and $z$ (below). Experimental 
data from Ref.~\cite{Seidl:2004dt}. Dashed line for the result of the spectator 
model at its scale $Q_0^2=0.3$ GeV$^2$, solid line for the result at the 
experimental scale $Q^2=2.5$ GeV$^2$ (see text for details about evolution).}
\label{fig:Autsinph+S}
\end{figure}

In Fig.~\ref{fig:Autsinph+S}, we show the $x$ (top panels) and $z$ (bottom panels) 
dependences of the weighted SSA $A_{UT}^{Q_T\,\sin (\phi_h+\phi_S)}$ from 
Eq.~(\ref{eq:Autsinph+S}) for both $\pi^+$ (left) and $\pi^-$ emission (right) 
in the same SIDIS kinematics of HERMES as in the previous figure (Collins 
effect~\cite{Collins:1993kk}). Experimental data are taken from 
Ref.~\cite{Seidl:2004dt}. The dashed line represents the result of the SSA when
calculating it at the model scale $Q_0^2=0.3$ GeV$^2$, where the analytic
expression of the Collins function $H_1^{\perp}$ is taken from the consistent 
spectator approach of Ref.~\cite{Bacchetta:2007wc}; the $D_1$ of 
Ref.~\cite{deFlorian:2007aj} has been down-evolved at LO using again the HOPPET
code~\cite{Salam:2008qg}. The solid line is the result for the SSA evolved at 
LO to the experimental scale $Q^2=2.5$ GeV$^2$. 

The agreement with data is satisfactory but for the unfavoured channel of 
$\pi^-$ emission, probably because of the approximation introduced into the
description of the Collins function for this case~\cite{Bacchetta:2007wc}. The 
effect of DGLAP evolution is not large, suggesting that there are compensations
between the numerator and denominator of the SSA.

\begin{figure}[h]
\begin{center}
\includegraphics[width=5.5cm]{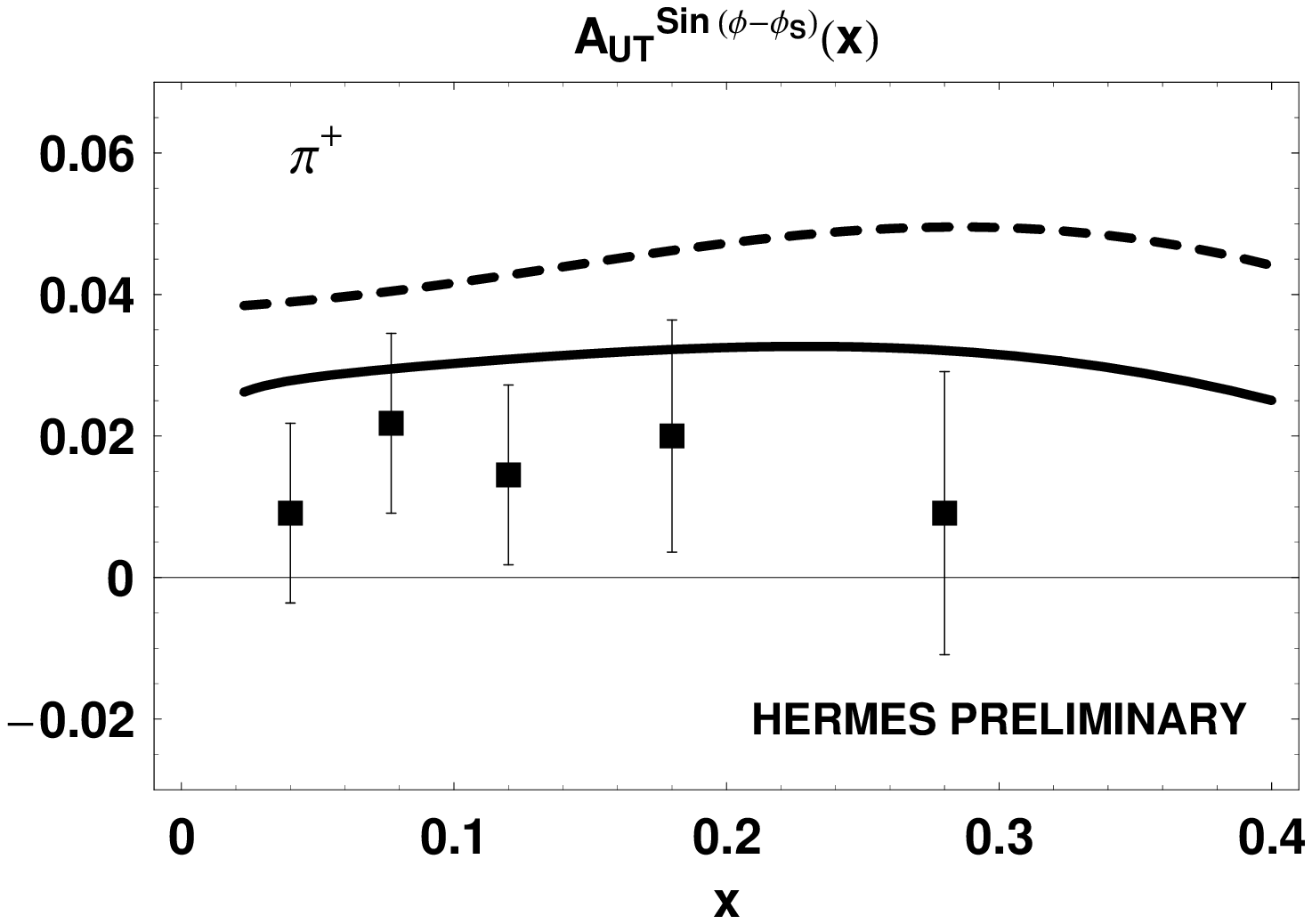} \hspace{0.2cm} 
\includegraphics[width=5.5cm]{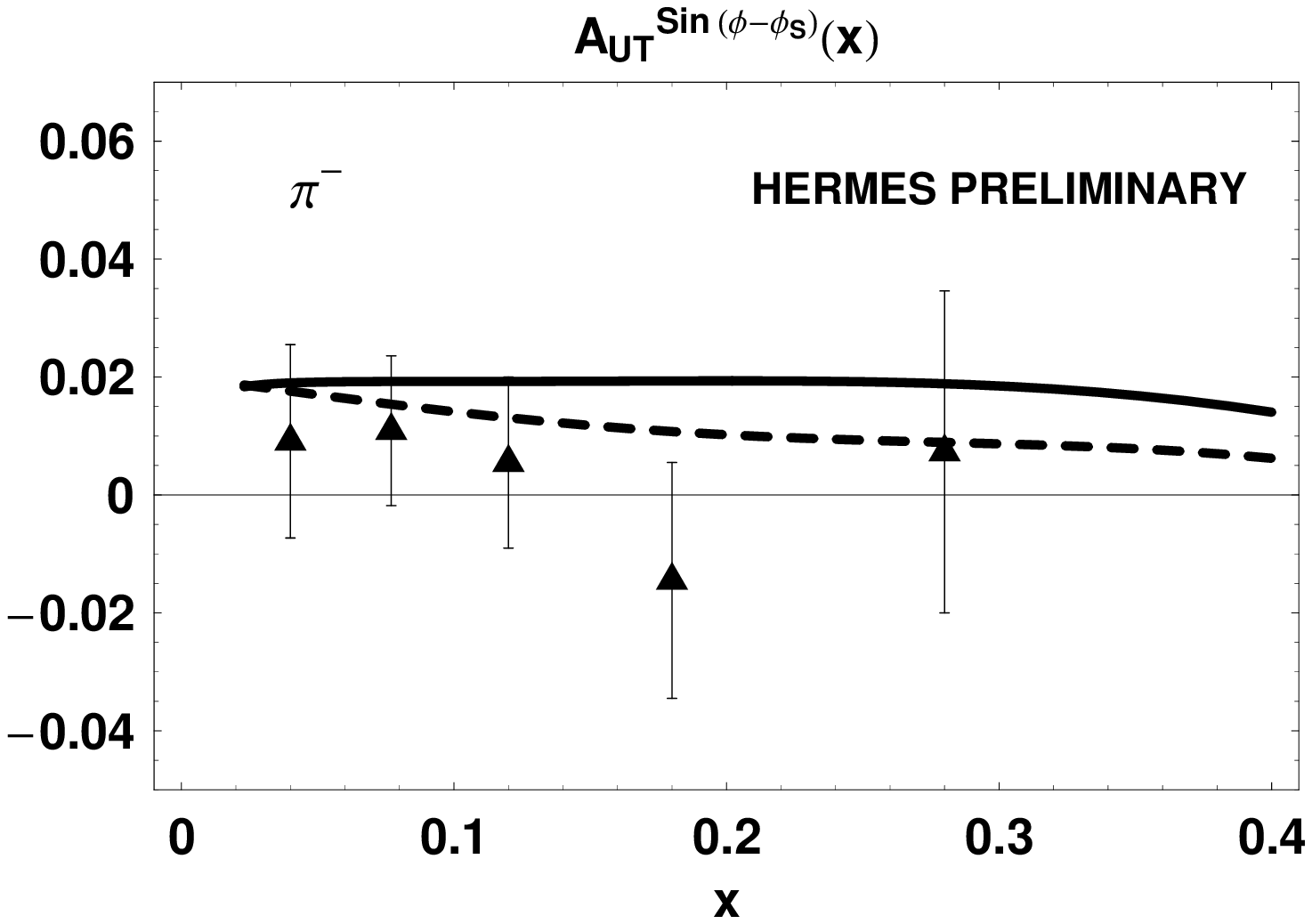} \\[0.2cm]
\includegraphics[width=5.5cm]{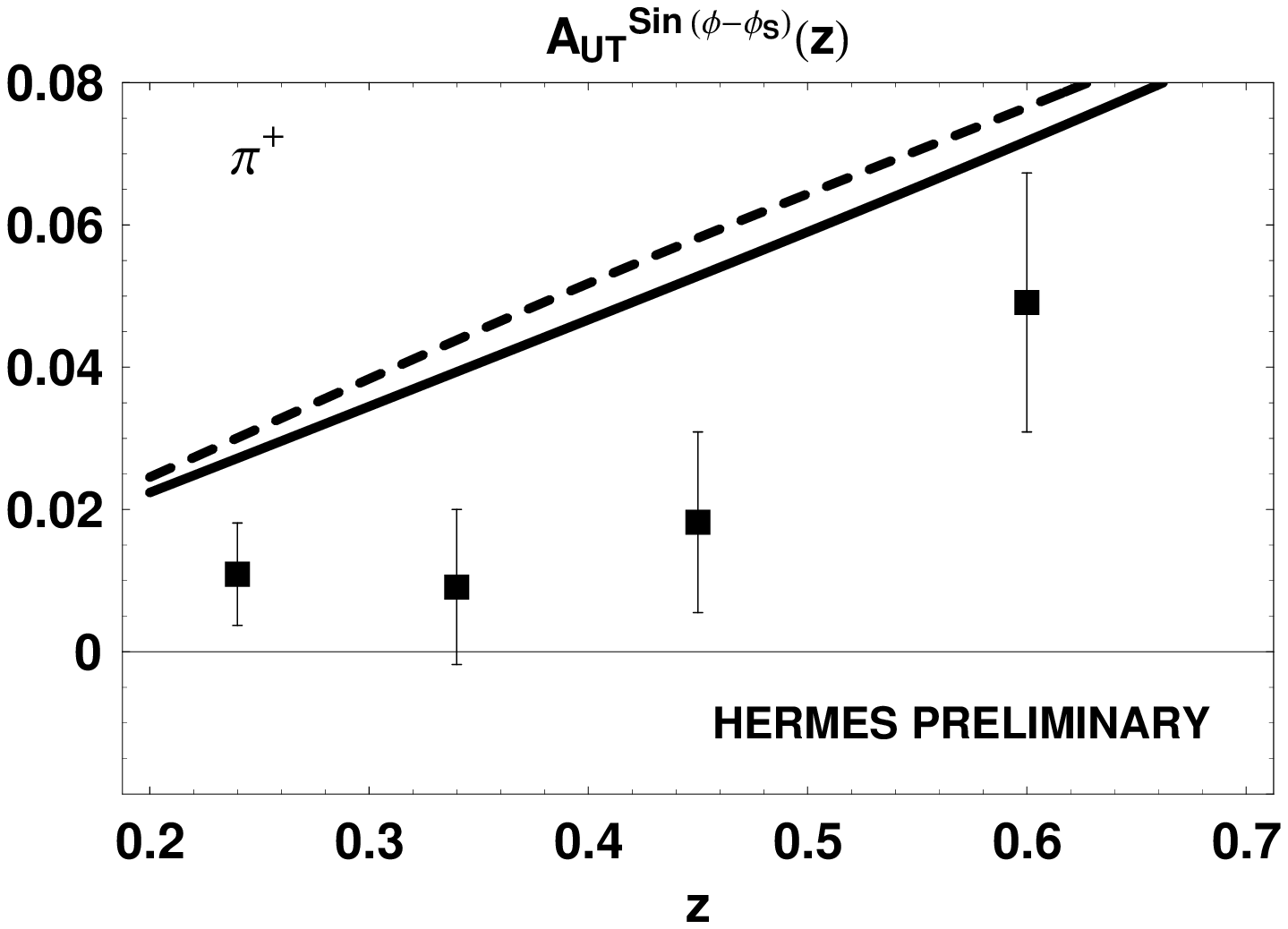} \hspace{0.2cm} 
\includegraphics[width=5.5cm]{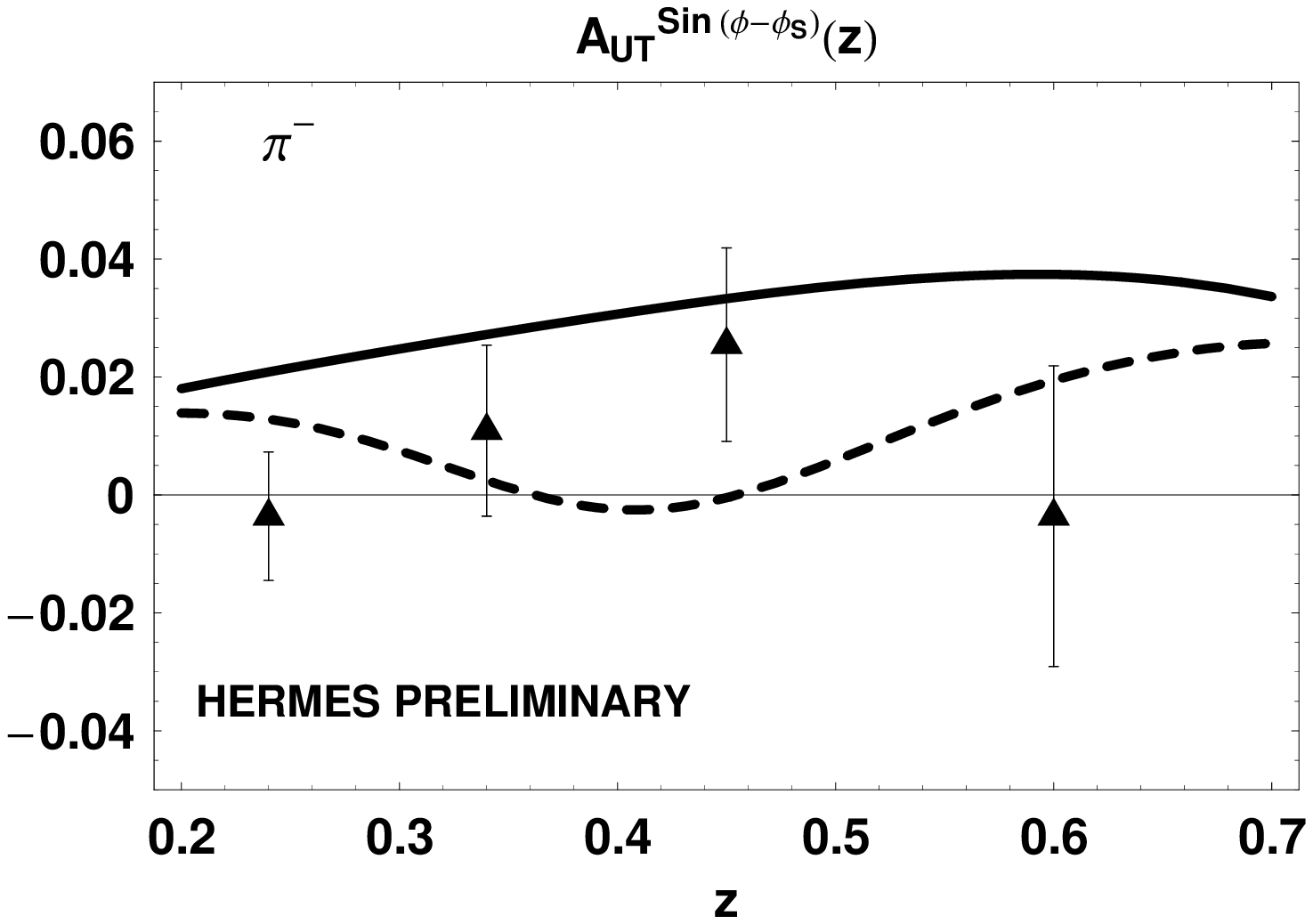}
\end{center}
\caption{The weighted single-spin asymmetry 
$A_{UT}^{Q_T\,\sin (\phi_h-\phi_S)}$ of Eq.~(\protect{\ref{eq:Autsinph-S}}) in 
the SIDIS kinematics of HERMES (Sivers effect~\cite{Sivers:1990cc}) with 
emission of $\pi^+$ (left) and $\pi^-$ (right), as a function of $x$ (above) 
and $z$ (below). Experimental data from Ref.~\cite{Seidl:2004dt}. Dashed and 
solid lines with the same notations as in previous figure.}
\label{fig:Autsinph-S}
\end{figure}

In Fig.~\ref{fig:Autsinph-S}, we show the $x$ (above) and $z$ (below) 
dependences of the weighted SSA $A_{UT}^{Q_T\,\sin (\phi_h-\phi_S)}$ from 
Eq.~(\ref{eq:Autsinph-S}) for both $\pi^+$ (left) and $\pi^-$ emission (right) 
in the same SIDIS kinematics of HERMES as in the previous figures (Sivers 
effect~\cite{Sivers:1990cc}). Experimental data are taken from 
Ref.~\cite{Seidl:2004dt}. The dashed line represents the result of the weighted 
SSA when calculating it at the model scale $Q_0^2=0.3$ GeV$^2$ by down-evolving 
the $D_1$ of Ref.~\cite{deFlorian:2007aj} at LO using again the HOPPET 
code~\cite{Salam:2008qg}. The solid line is the result for the weighted SSA 
evolved at LO to the experimental scale $Q^2=2.5$ GeV$^2$. 

The agreement between our results and the data is good for the $x$ dependence, 
while a discrepancy is evident for the $z$ distribution in the $\pi^+$ channel. 
As already anticipated at the beginning of Sec.~\ref{sec:out}, the lacking of a strong 
sea--quark contribution at small $x$ (in our model, it is generated only by radiative 
corrections) may deplete the denominator of the SSA once integrated upon $x$, and 
produce the observed enhancement upon data in the $z$ distribution (in the numerator, 
the Sivers function is less affected by sea quarks).

\begin{figure}[h]
\begin{center}
\includegraphics[width=6cm]{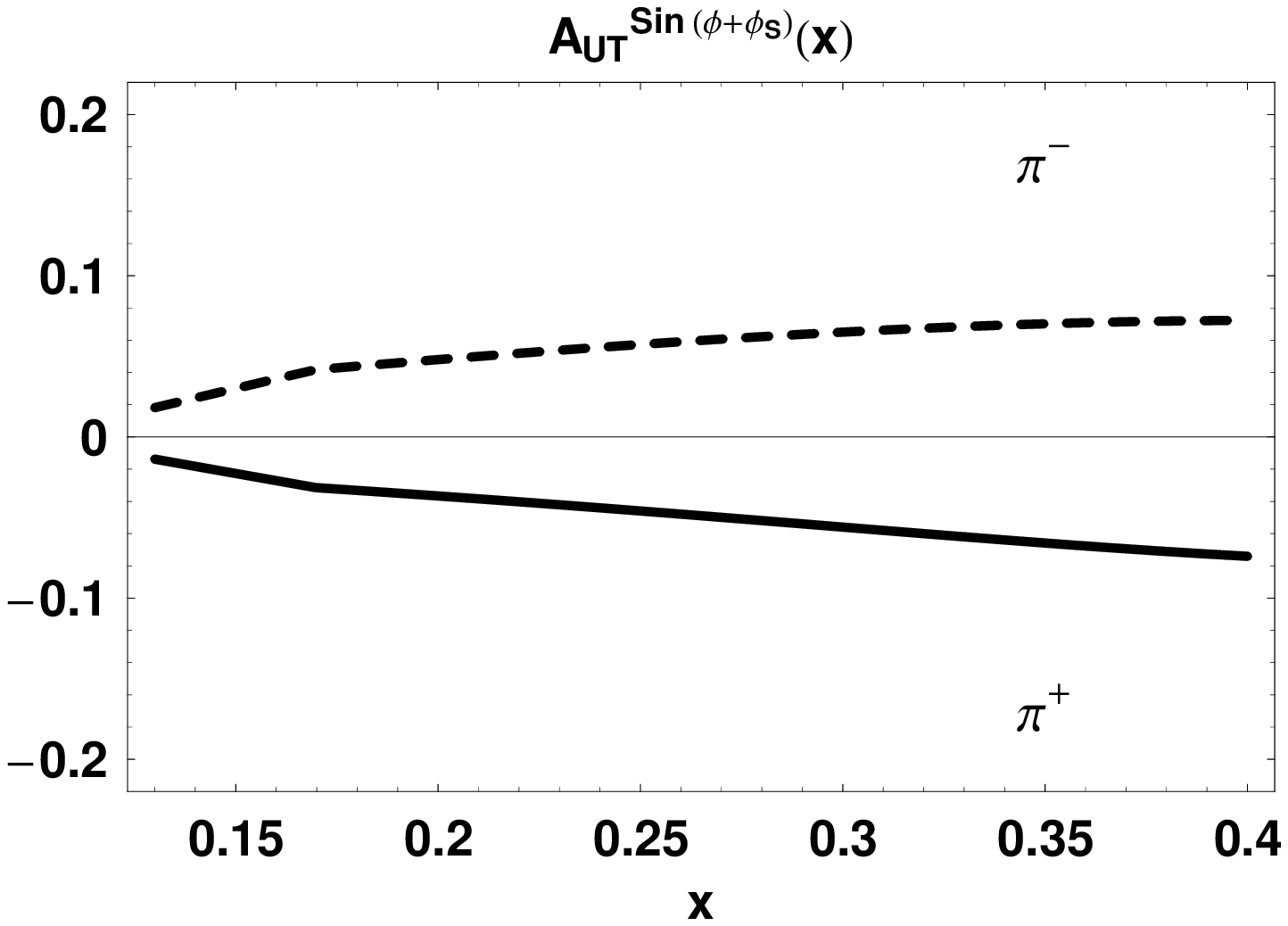} \hspace{0.2cm} 
\includegraphics[width=6cm]{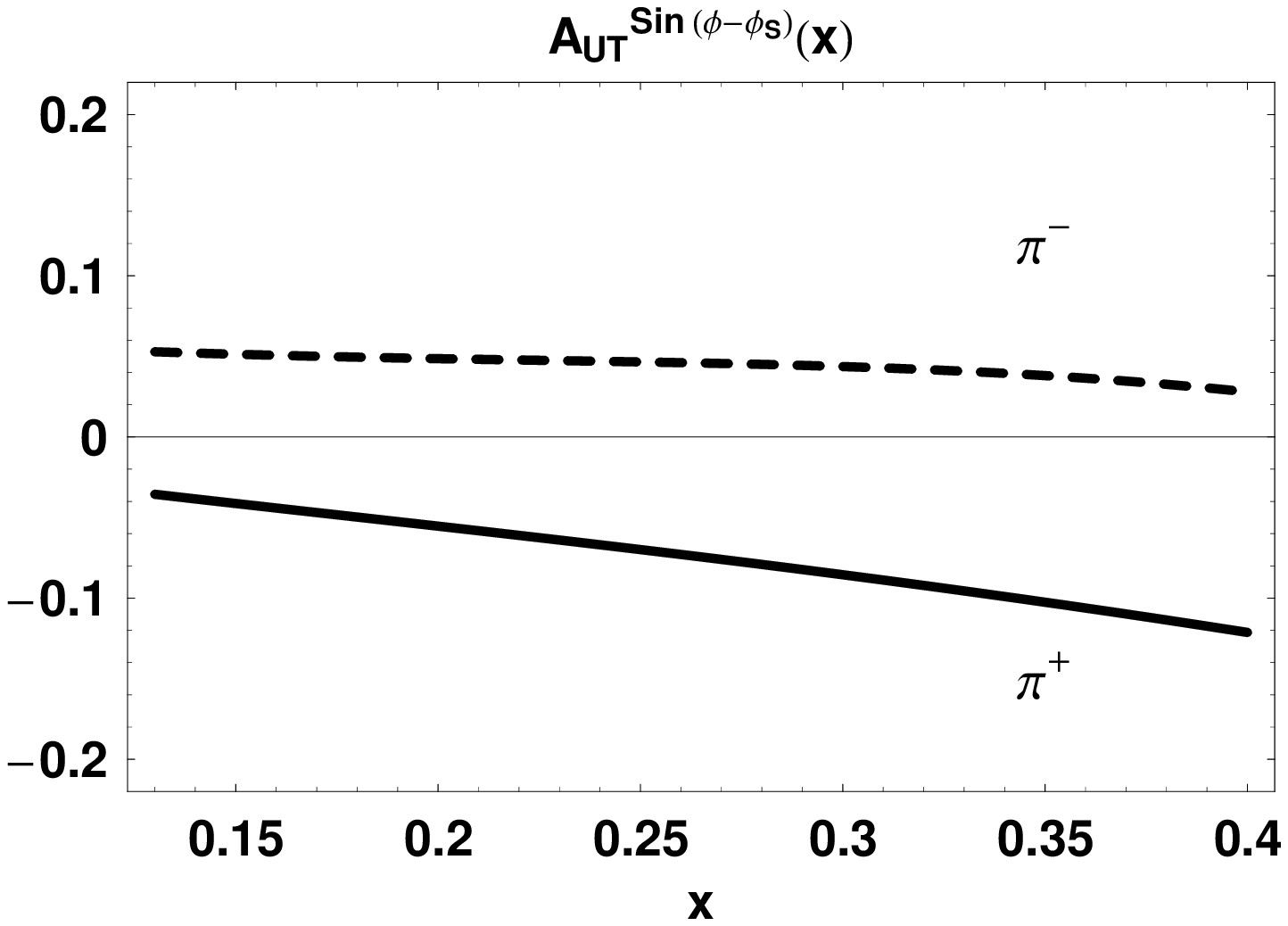} \\[0.2cm]
\includegraphics[width=6cm]{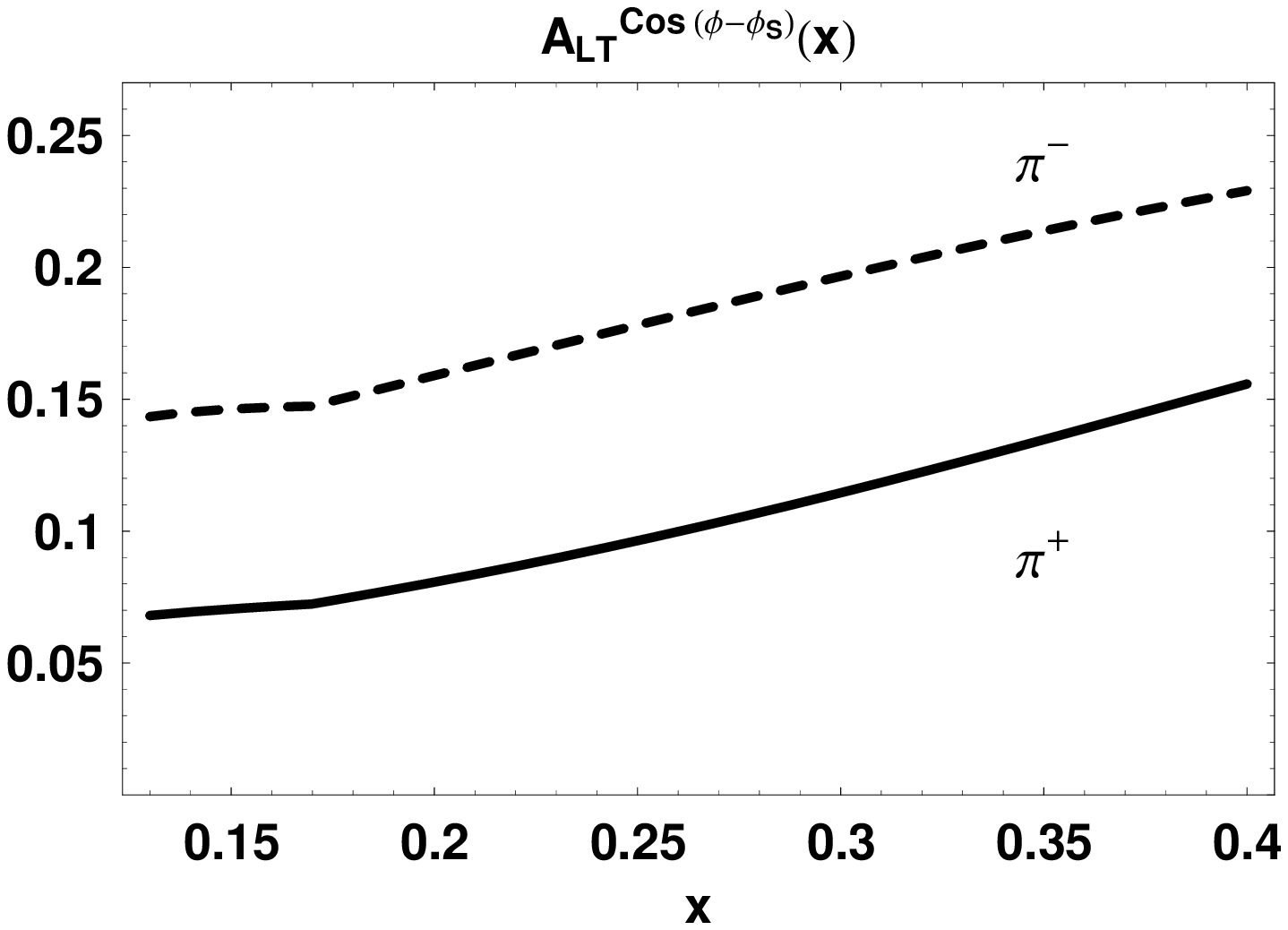}
\end{center}
\caption{The weighted single-spin asymmetries 
$A_{UT}^{Q_T\,\sin (\phi_h+\phi_S)}$ of Eq.~(\protect{\ref{eq:Autsinph+S}}) 
(Collins effect), $A_{UT}^{Q_T\,\sin (\phi_h-\phi_S)}$ of 
Eq.~(\protect{\ref{eq:Autsinph-S}}) (Sivers effect), 
$A_{LT}^{Q_T\,\cos (\phi_h-\phi_S)}$ of Eq.~(\protect{\ref{eq:Altcosph-S}}), as 
functions of $x$ in the SIDIS kinematics at JLab with a 6 GeV beam energy for
the emission of $\pi^+$ (solid line) and $\pi^-$ (dashed line) from a
transversely polarized neutron~\cite{e06-010}.}
\label{fig:JLab}
\end{figure}

In Fig.~\ref{fig:JLab}, we show our predictions for the $x$ dependence of 
the weighted SSA $A_{UT}^{Q_T\,\sin (\phi_h+\phi_S)}$ of 
Eq.~(\ref{eq:Autsinph+S}) (Collins effect), 
$A_{UT}^{Q_T\,\sin (\phi_h-\phi_S)}$ of Eq.~(\ref{eq:Autsinph-S}) (Sivers
effect), and $A_{LT}^{Q_T\,\cos (\phi_h-\phi_S)}$ of Eq.~(\ref{eq:Altcosph-S}), 
for the reaction $n^\uparrow (e,e' \pi^\pm)$. The measurement was recently
performed at Hall A of JLab by the E06-010 and E06-011 
collaborations~\cite{e06-010}, using a 6 GeV energy beam and a transversely
polarized $^3\mathrm{He}$ (effective neutron) target in SIDIS kinematics with cm 
energy squared $s=12.14$ GeV$^2$, average scale $\langle Q^2\rangle = 2.2$ 
GeV$^2$ and the following experimental cuts: 
\begin{align}
& 0.13<x<0.4 \; , &0.46<z<0.59 \; , \hspace{5.5cm}  \nn \\
& y_{\text{min}}(x)<y<0.86 \; , &y_{\text{min}}(x)=\text{max} \left[ 0.68,\, 
\frac{1.3}{x (s-M^2)}, \, \frac{5.4-M^2}{(1-x)\, (s-M^2)} \right] \; . 
\label{eq:sidiskin1}
\end{align}

In each panel of Fig.~\ref{fig:JLab}, the solid (dashed) line represents the 
result for the emission of $\pi^+$ ($\pi^-$) in the above kinematics. All 
components of the weighted SSA have been evolved at LO to the experimental 
average scale $Q^2=2.2$ GeV$^2$ using the HOPPET code~\cite{Salam:2008qg}. The 
LO evolution kernels for the transversity and the first $\bm{p}_{\sT}$ moment 
of the Sivers function are the same as those ones used in 
Fig.~\ref{fig:Autsinph+S} and~\ref{fig:Autsinph-S}, respectively. The evolution 
of $g_{1T}^{(1)}$ is assumed at LO to be the same as the one of the helicity 
distribution $g_1$.


\subsection{The SSA in Drell--Yan }
\label{sec:dy-out}

As there are no data for weighted SSA in Drell--Yan collisions, we will show 
our predictions for cases of interest in view of future experiments. At FAIR
(GSI), the PAX collaboration is planning to measure the fully polarized 
Drell--Yan process with antiprotons in order to perform a self-consistent 
extraction of the transversity distribution~\cite{Barone:2005pu,Maggiora:2005cr,Efremov:2004qs,Anselmino:2004ki,Bianconi:2005bd}.
 Moreover, by simply switching on and off the transverse polarization of only
one hadron, from the combination of the various cross sections it is possible 
to extract $h_1$ also together with the Boer--Mulders function $h_1^{\perp}$
[see Eqs.~(\ref{eq:dyAutsinph-S}, \ref{eq:dyAutsinph+S})]. Finally, using the 
same process the COMPASS collaboration is planning to extract the Sivers
function using a high energetic pion beam on a transversely polarized proton 
target at CERN~\cite{Efremov:2004tp,Bianconi:2006hc,Bianconi:2005yj,Anselmino:2009st}, 
in order to directly test its predicted non-universal 
behaviour~\cite{Collins:2002kn}.

\begin{figure}[h]
\begin{center}
\includegraphics[width=7.5cm]{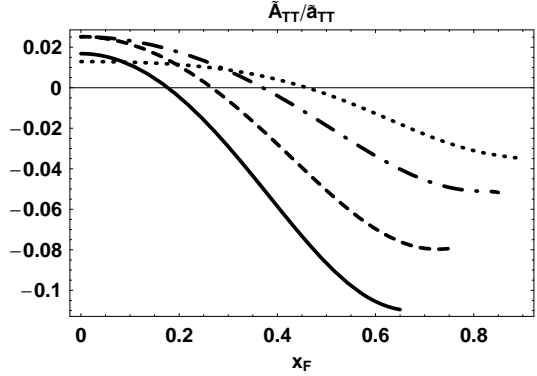} 
\end{center}
\caption{The double-spin asymmetry $\tilde{A}_{TT}/\tilde{a}_{TT}$ of 
Eq.~(\protect{\ref{eq:dyatt}}) as a function of $x_F$ for the 
$\bar{p}^\uparrow p^\uparrow \rightarrow \mu^+ \mu^- X$ process at the PAX 
kinematics (see text). Solid, dashed, dot-dashed, and dotted lines correspond to
c.m. energies squared $s=30,\,45,\,80,\,200$ GeV$^2$, respectively.}
\label{fig:dyatt}
\end{figure}

In Fig.~\ref{fig:dyatt}, we show our prediction for the $x_F$ dependence 
of the double-spin asymmetry $\tilde{A}_{TT}/\tilde{a}_{TT}$ from 
Eq.~(\ref{eq:dyatt}) for the 
$\bar{p}^\uparrow p^\uparrow \rightarrow \mu^+ \mu^- X$ process that could be 
studied at PAX with a fully transversely polarized antiproton beam 
($|\bm{S}_{1\sT}|=1$) colliding on a fully ($|\bm{S}_{2\sT}|=1$) transversely 
polarized proton target~\cite{Efremov:2004qs,Barone:2005pu}. Conventions are 
according to Eq.~(\ref{eq:dykinvar}). Dimuon invariant masses are summed in the 
range $2<M<3$ GeV, below the $J/\psi$ resonance. Solid, dashed, dot-dashed, and 
dotted lines correspond to c.m. energies squared $s=30,\,45,\,80,\,200$ 
GeV$^2$, respectively. All the parton distributions entering the asymmetry are 
evolved at LO to each $Q^2\equiv M^2$ inside the integration range. 

Despite the fact that $\tilde{A}_{TT}$ is roughly the ``squared" of the
transversity distribution, it does change sign for some $x_F$ depending on the
value of $s$. As it is evident from Eq.~(\ref{eq:dykinvar}), for different $s$ 
the same $x_F$ and $M$ probe different $x_1$ and $x_2$, and, in particular, 
those where one of the two model transversities for quark up changes sign (see 
Fig.~8 of Ref.~\cite{Bacchetta:2008af}). Consequently, the asymmetry of 
Eq.~(\ref{eq:dyatt}) also changes sign. This feature contrasts with other 
results in the literature~\cite{Efremov:2004qs,Anselmino:2004ki,Barone:2005cr,Pasquini:2006iv}. 
Moreover, we also see a monotonically decreasing trend in $x_F$ (and also in 
$y$), but our asymmetry is as large as 10\% in modulus at most. This is 
probably due to cancellations that occur when summing upon the invariant mass 
$M$, or equivalently when integrating upon specific portions of the $(x_1,x_2)$ 
phase space, where the product $h_1^{\overline{u}}(x_1)\, h_1^u(x_2)$ in 
$\tilde{A}_{TT}$ repeatedly changes sign [and is comparable with, or bigger 
than, the positive $h_1^{\overline{d}}(x_1)\, h_1^d(x_2)$]. For both reasons, a 
measure of $\tilde{A}_{TT}$ is highly desirable because it would also clarify 
the issue.

\begin{figure}[h]
\begin{center}
\includegraphics[width=7.5cm]{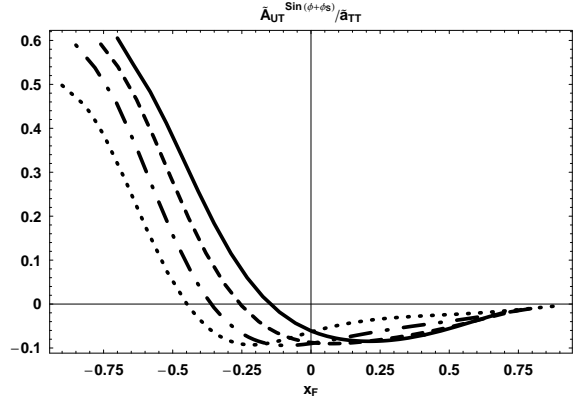} 
\end{center}
\caption{The weighted single-spin asymmetry 
$\tilde{A}_{UT}^{q_T\,\sin (\phi +\phi_{S_2})}/\tilde{a}_{TT}$ of 
Eq.~(\protect{\ref{eq:dyAutsinph+S}}) for the 
$\bar{p} p^\uparrow \rightarrow \mu^+ \mu^- X$ process as a function of $x_F$, 
in the same kinematics, conventions and notations, as in the previous figure.}
\label{fig:dyAutsinph+S}
\end{figure}

In Fig.~\ref{fig:dyAutsinph+S}, we show our prediction for the $x_F$ dependence 
of the weighted SSA 
$\tilde{A}_{UT}^{q_T\,\sin (\phi +\phi_{S_2})}/\tilde{a}_{TT}$ from 
Eq.~(\ref{eq:dyAutsinph+S}) for the 
$\bar{p} p^\uparrow \rightarrow \mu^+ \mu^- X$ process that could be studied at 
PAX with antiproton beams colliding on a transversely polarized proton
target~\cite{Bianconi:2004wu,Barone:2005pu,Radici:2007vc}. Conventions and 
notations are the same as in the previous figure. Evolution effects are included
for all partonic densities; as already explained at the end of 
Sec.~\ref{sec:dy}, the LO evolution equations of the chiral-odd 
$h_1^{\perp (1)}$ are assumed to be the same as those of the chiral-odd
transversity $h_1$. The outcome in Fig.~\ref{fig:dyAutsinph+S} already takes
into account the predicted sign change of T-odd TMDs when going from SIDIS to
Drell--Yan collisions~\cite{Collins:2002kn}. 

Again, for different $s$ the same $x_F$ and $M$ probe different $x_1$ and $x_2$,
in particular the range where the $h_1^u(x_2)$ changes sign. Consequently, the 
asymmetry of Eq.~(\ref{eq:dyAutsinph+S}) also changes sign, the turning point 
depending on $s$. For the very same reason, cancellations probably occur when 
summing upon the invariant mass $M$, which suppress the size of the asymmetry at
positive $x_F$. On the contrary, the size of 
$\tilde{A}_{UT}^{q_T\,\sin (\phi +\phi_{S_2})}/\tilde{a}_{TT}$ is significant at
negative $x_F$. .

\begin{figure}[h]
\begin{center}
\includegraphics[width=7cm]{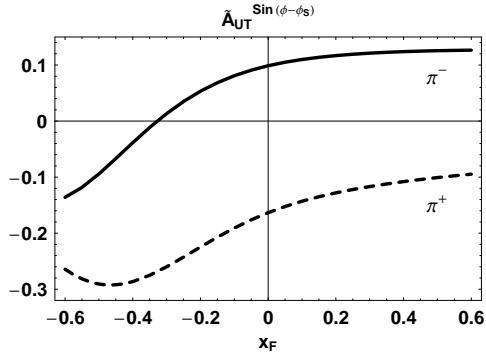} 
\end{center}
\caption{The weighted single-spin asymmetry 
$\tilde{A}_{UT}^{q_T\,\sin (\phi -\phi_{S_2})}$ of 
Eq.~(\protect{\ref{eq:dyAutsinph-S}}) as a function of $x_F$ for the 
$\pi p^\uparrow \rightarrow \mu^+ \mu^- X$ process at COMPASS (see text). Upper 
solid (lower dashed) line for $\pi^-$ ($\pi^+$).}
\label{fig:dyAutsinph-S}
\end{figure}

In Fig.~\ref{fig:dyAutsinph-S}, we show our prediction for the $x_F$ dependence 
of the weighted SSA $\tilde{A}_{UT}^{q_T\,\sin (\phi -\phi_{S_2})}$ from 
Eq.~(\ref{eq:dyAutsinph-S}) for the $\pi p^\uparrow \rightarrow \mu^+ \mu^- X$ 
process that could be studied at COMPASS with a 160 GeV pion beam colliding on 
a transversely polarized proton 
target~\cite{Efremov:2004tp,Bianconi:2006hc,Anselmino:2009st}. Dimuon invariant 
masses are summed in the safe range $4<M<9$ GeV between the $J/\psi$ and the 
$\Upsilon$ resonances; since $s=300$ GeV$^2$, the invariant $\tau \approx 0.16$ 
is in the valence region. Other kinematic onventions are the same as in the 
previous figure. Again, the result already takes into account the predicted 
sign change of T-odd TMDs when going from SIDIS to Drell--Yan 
collisions~\cite{Collins:2002kn}. 

Following Ref.~\cite{Efremov:2004tp}, we approximate the 
$\tilde{A}_{UT}^{q_T\,\sin (\phi -\phi_{S_2})}$ of Eq.~(\ref{eq:dyAutsinph-S})
by considering only the dominant valence contribution to the $\pi-p$ collision. 
Therefore, for $\pi^-$ ($\pi^+$) we retain only the $\bar{u}-u$ ($\bar{d}-d$) 
annihilation and the final weighted SSA does no longer depend on the partonic
densities in the pion. Evolution effects are included at LO along the lines 
described at the end of Sec.~\ref{sec:sidis}. The upper solid (lower dashed) 
line is our result for $\pi^-$ ($\pi^+$). 

The size of the asymmetry should allow for an unambiguous test of the prediction
about the above mentioned sign change of the Sivers function with respect to 
its extraction from SIDIS~\cite{Bianconi:2006hc}.


\section{Conclusions}
\label{sec:end}

In this paper, we have analytically calculated azimuthal (spin) asymmetries in 
lepton-nucleon semi-inclusive deep-inelastic scattering (SIDIS) and in Drell-Yan 
hadronic collisions, using the leading-twist transverse-momentum dependent 
distributions (TMDs) obtained in the diquark spectator model of the nucleon from 
Ref.~\cite{Bacchetta:2008af}. 

We have included evolution effects at leading order (LO) in $\alpha_s$ by 
implementing in the HOPPET code of Ref.~\cite{Salam:2008qg} the kernel for 
chiral-odd objects. While DGLAP equations are well known for the momentum 
$f_1(x)$, helicity $g_1(x)$, and transversity $h_1(x)$ distributions,  we have 
included evolution effects in an approximate way also for the first 
$\bm{p}_{\sT}$ moments of the Sivers $f_{1T}^{\perp\, (1)}(x)$, $g_{1T}^{(1)}(x)$, 
and Boer-Mulders $h_1^{\perp\, (1)}(x)$ functions, as described in 
Sec.~\ref{sec:out}. Moreover, since in our model the first $\bm{p}_{\sT}$ moment 
of na\"ive T-odd densities depends linearly upon $\alpha_s$ at the model scale, 
a consistent treatment of evolution effects requires to determine $\alpha_s$ itself 
from the renormalization group equations. Therefore, with respect to 
Ref.~\cite{Bacchetta:2008af} we have modified its value at the hadronic model 
scale $Q_0^2$: instead of using an {\it ad-hoc} nonperturbative input as a free 
parameter, we have computed $\alpha_s (Q_0^2)$ from the LO renormalization 
group equations. The net effect is a resizing of all T-odd functions by a global constant. 
The comparison with available 
parametrizations~\cite{Anselmino:2008sg,Lu:2009ip,Barone:2009hw} is encouraging 
for the Sivers function, but requires more investigations in the case of the Boer-Mulders 
one. 

We have tested our model TMDs by considering weighted single-spin asymmetries 
(SSA) in both SIDIS and Drell--Yan collisions. Data for weighted SSA are scarce, with 
low statistics, and still preliminary~\cite{Airapetian:2004zf,Seidl:2004dt}. But, from 
the theoretical side they are preferable than unweighted SSA because in a model 
independent way their final expressions get automatically factorized involving simple 
products of ``collinear" objects~\cite{Mulders:1995dh} -- parton distribution functions (PDFs) or 
$\bm{p}_{\sT}$ moments of TMDs (and analogously for fragmentation functions). 
As such, weighted SSA can be always calculated analytically in our model. Incidentally, 
the choice of considering evolution effects only at LO is dictated by the consistency 
with the expressions for the weighted SSA, which are known at LO. 

As for SIDIS, we have considered the asymmetry $A_{LL}$ with longitudinally 
polarized protons~\cite{Airapetian:2004zf}, and the Collins and the Sivers 
effects~\cite{Seidl:2004dt}, all measured by the HERMES collaboration; we 
have made predictions for the Collins and the Sivers effects, as well as for 
$A_{LT}$, in the kinematic conditions recently explored in the E06-010 and 
E06-011 experiments in Hall A at JLab~\cite{e06-010}, using the 6 GeV energy 
beam and the transversely polarized $^3\mathrm{He}$ effective neutron target. 
For the unpolarized fragmentation function $D_1$, we have adopted the 
parametrization of Ref.~\cite{deFlorian:2007aj}, while we have used the 
analytic expression of the Collins function from Ref.~\cite{Bacchetta:2007wc}, 
obtained using a spectator approach similar to the present framework. 
Overall, the comparison with experimental data is satisfactory. For the Collins 
effect, in some cases there are discrepancies that probably can be traced back 
to the assumptions made in the description of the unfavoured channels of the 
Collins function~\cite{Bacchetta:2007wc}. 

Since there are no data for weighted SSA in Drell--Yan collisions, we have made 
predictions for cases of interest in view of future experiments. At FAIR
(GSI), the PAX collaboration is planning to measure the fully polarized 
Drell--Yan process with antiprotons in order to perform a self-consistent 
extraction of the transversity 
distribution~\cite{Barone:2005pu,Maggiora:2005cr,Efremov:2004qs,Anselmino:2004ki,Bianconi:2005bd};
we have presented our predictions for $\tilde{A}_{TT}$ in various kinematic
configurations. Interestingly, despite the fact that this asymmetry is
approximately the ``squared" of the transversity distribution $h_1$, our 
$\tilde{A}_{TT}$ does change sign in some portions of the phase space. Briefly,
the kinematics of the parton-antiparton annihilation explores different ranges 
in the $x$ dependence of the two involved $h_1$, which can separately become
negative in our model (see Fig.~8 of Ref.~\cite{Bacchetta:2008af}). This feature contrasts 
with other results in the 
literature~\cite{Efremov:2004qs,Anselmino:2004ki,Barone:2005cr,Pasquini:2006iv};
it would be highly desirable to measure $\tilde{A}_{TT}$ in order to clarify 
this issue.  

By simultaneously considering Drell--Yan collisions where both hadrons are unpolarized or 
only one is transversely polarized, it is possible to measure a weighted SSA that leads to 
the extraction of $h_1$ in combination with the Boer-Mulders function 
$h_1^{\perp}$. We have presented our predictions again in the kinematic regime
that the PAX collaboration could explore at FAIR (GSI). The same previous
comment on the sign of $h_1$ applies here too. 

Finally, using the same process (single-polarized Drell--Yan collision) the 
COMPASS collaboration is planning to extract the Sivers function using a high 
energetic pion beam on a transversely polarized proton 
target at CERN~\cite{Efremov:2004tp,Bianconi:2006hc,Bianconi:2005yj,Anselmino:2009st}, 
in order to directly test its predicted non-universal 
behaviour~\cite{Collins:2002kn}. We have shown our predictions for both $\pi^-$
and $\pi^+$ collisions, assuming that the elementary annihilation is driven by
the dominant valence contributions. 


\section*{Acknowledgments}

This work was partially supported by the Research Infrastructure Integrated Activity ``Study 
of Strongly Interacting Matter" (acronym HadronPhysics2, grant agreement n. 227431) under 
the $7^{th}$ Framework Programme of the European Community, and by the Italian MIUR 
through the PRIN 2008EKLACK ``Structure of the nucleon: transverse momentum, transverse 
spin and orbital angular momentum".


\bibliographystyle{apsrev}
\bibliography{alebiblio}


\end{document}